\documentclass[twocolumn,english,onecolumn]{IEEEtran}
\usepackage[T1]{fontenc}
\usepackage{babel}
\usepackage{array}
\usepackage{prettyref}
\usepackage{multirow}
\usepackage{amsthm}
\usepackage{amsmath}
\usepackage{amssymb}
\usepackage{graphicx}
\usepackage[unicode=true,
 bookmarks=true,bookmarksnumbered=true,bookmarksopen=true,bookmarksopenlevel=1,
 breaklinks=false,pdfborder={0 0 0},backref=false,colorlinks=false]
 {hyperref}
\hypersetup{pdftitle={Your Title},
 pdfauthor={Your Name},
 pdfpagelayout=OneColumn, pdfnewwindow=true, pdfstartview=XYZ, plainpages=false}
\usepackage{breakurl}

\makeatletter

\providecommand{\tabularnewline}{\\}

\theoremstyle{plain}
\newtheorem{thm}{\protect\theoremname}
\theoremstyle{plain}
\newtheorem{prop}[thm]{\protect\propositionname}

\usepackage{babel}
\ifCLASSOPTIONcompsoc
\else
\fi
\usepackage{cite}\usepackage{MnSymbol}\usepackage{subfigure}

\usepackage{balance}

\usepackage{enumitem}

\usepackage{babel}

\providecommand{\theoremname}{Theorem}

\makeatother

\providecommand{\propositionname}{Proposition}
\providecommand{\theoremname}{Theorem}

\begin{document}

\title{Reliable Linear, Sesquilinear and Bijective Operations On Integer
Data Streams Via Numerical Entanglement}

\author{Mohammad Ashraful Anam and~Yiannis Andreopoulos\emph{$^{*}$, Senior
Member, IEEE}%
\thanks{$^{*}$Corresponding author. This paper will appear in the IEEE Trans. on Signal Processing. This work was supported by EPSRC, project
EP/M00113X/1. The authors are with the Electronic and Electrical Engineering
Department, University College London, Roberts Building, Torrington
Place, London, WC1E 7JE, Tel. +44 20 7679 7303, Fax. +44 20 7388 9325
(both authors), Email: \{mohammad.anam.10, i.andreopoulos\}@ucl.ac.uk.
Parts of this paper have been presented at the 2015 IEEE International
On-Line Testing Symposium (IEEE IOLTS 2015). %
}}
\maketitle
\begin{abstract}
A new technique is proposed for fault-tolerant linear, sesquilinear
and bijective (LSB) operations on $M$ integer data streams ($M\geq3$),
such as: scaling, additions/subtractions, inner or outer vector products,
permutations and convolutions. In the proposed method, the $M$ input
integer data streams are linearly superimposed to form $M$ \emph{numerically-entangled}
integer data streams that are stored in-place of the original inputs.
A series of LSB operations can then be performed directly using these
entangled data streams. The results are extracted from the $M$ entangled
output streams by additions and arithmetic shifts. Any soft errors
affecting any single disentangled output stream are guaranteed to
be detectable via a specific post-computation reliability check. In
addition, when utilizing a separate processor core for each of the
$M$ streams, the proposed approach can recover all outputs after
any single fail-stop failure. Importantly, unlike algorithm-based
fault tolerance (ABFT) methods, the number of operations required
for the entanglement, extraction and validation of the results is
linearly related to the number of the inputs and does not depend on
the complexity of the performed LSB operations. We have validated
our proposal in an Intel processor (Haswell architecture with AVX2
support) via several types of operations: fast Fourier transforms,
circular convolutions, and matrix multiplication operations. Our analysis
and experiments reveal that the proposed approach incurs between $0.03\%$
to $7\%$ reduction in processing throughput for a wide variety of
LSB operations. This overhead is 5 to 1000 times smaller than that
of the equivalent ABFT method that uses a checksum stream. Thus, our
proposal can be used in fault-generating processor hardware or safety-critical
applications, where high reliability is required without the cost
of ABFT or modular redundancy. \end{abstract}


\section{Introduction}

\IEEEPARstart{T}{he} current technology roadmap for high-performance
computing (HPC) indicates that digital signal processing (DSP) routines
running on such systems must become resilient to transient or permanent
faults occurring in arithmetic, memory or logic units. Such faults
can be caused by process variations, silent data corruptions (e.g.,
due to particle strikes, circuit overclocking or undervolting, or
other hardware non-idealities) \cite{nicolaidis2012design}, scheduling
and runtime-induced faults \cite{AndreopoulosTMM_PlentyOfRoom}, misconfigured
application programming interfaces (APIs) \cite{lu2013cloud} and
opportunistic resource reservation \cite{poola2014fault} in cloud
computing systems. For example, DSP systems such as: webpage or multimedia
retrieval \cite{carterette2009million}, object or face recognition
in images \cite{yang2004two}, machine learning and security applications
\cite{bradski2008learning}, transform decompositions and video encoding
\cite{andreopoulos2000hybrid,andreopoulos2002new,andreopoulos2001local,munteanu2003control,andreopoulos2003high,andreopoulos2007adaptive,kontorinis2009statistical,foo2008analytical,barbarien2004scalable},
and visual search and retrieval systems \cite{jegou2012aggregating},
are now deployed using Amazon Elastic Compute Cloud (EC2) spot instances
with substantially-reduced billing cost (e.g., in the order of 0.01\$
per core per hour). However, Amazon reserves the right to terminate
EC2 spot instances at any moment with little or no prior notice. In
addition, transient service interruptions may occur at unpredictable
intervals, since processor cores in EC2 spot instance reservations
may not be solely dedicated to the cluster under consideration \cite{lu2013cloud,poola2014fault}. 

System-induced faults in DSP routines manifest as \cite{nicolaidis2012design,AndreopoulosTMM_PlentyOfRoom,fiala2012detection}:
\emph{(i)} \emph{transient faults}, where execution continues uninterrupted
on all input data streams---albeit with corrupted data and possibly
carrying out erroneous logic or arithmetic operations---or \emph{(ii)}
\emph{fail-stop failures}, where the execution on one of the processor
cores halts due to a fail-stop exception (e.g., overflow detection,
memory leak assertion, etc.) or a system crash. In the first case,
a highly-reliable system should be able to detect all faults (and
possibly correct them); in the second case (which is analogous to
an \emph{erasure} in a communications system, since it is easily detectable),
the system should be able to recover the results of the halted execution
without requiring recomputation. 

The compute- and memory-intensive parts within DSP routines comprise:
transform decompositions, signal cross-correlation, inner or outer
vector products, matrix products, etc. Such operations are linear,
sesquilinear (also known as ``one-and-half linear'') and bijective,
collectively called LSB operations in this paper. Moreover, especially
when considering multimedia data, these operations are typically performed
using 32-bit or 64-bit integer arithmetic. Therefore, ensuring \emph{highly
reliable} integer LSB operations with minimal overhead against their
\emph{unreliable} equivalents is of paramount importance for signal
processing systems.

\subsection{Summary of Prior Work }

Existing techniques that can ensure reliability to transient faults
and/or fail-stop failures comprise three categories: \emph{(i)} component-based
reliability via error-correcting codes (ECC) \cite{alameldeen2011energy},
e.g., random access memory or cache memory chip designs with native
ECC support for the detection/correction of bit flips;\emph{ (ii)}
algorithm-based fault-tolerance (ABFT) \cite{huang1984algorithm,chen2005fault,luk1985weighted,stefanidis2004weighted,nair1988linearCode,sloan2012algorithmic,rexford1992partitioned},
i.e., methods producing additional checksum inputs/outputs that are
tailored to the algorithm under consideration; \emph{(iii)} systems
with double or triple modular redundancy (MR), where the same operation
is performed in parallel in two or three separate processors (or threads)
that cross-validate their results and recover from fail-stop failures
\cite{engelmann2009case}. 

While component-based methods can indeed mitigate many of the faults
occurring in memory or processor components, their integration into
hardware designs is known to incur substantial overhead \cite{fatica2009accelerating}.
Therefore, system designers tend to disable such functionalities for
most applications, even at the risk of allowing faults to occur undetected.
Moreover, in order to offer end-to-end guarantees on fault tolerance
and robustness to fail-stop failures (i.e., from the input data streams
to the results of a certain computation), component-based fault tolerance
must be integrated with checkpointing and execution roll-back \cite{chen2005fault},
which further increases the complexity of the final solution. 

Similarly, it is well known that ABFT and MR solutions can lead to
substantial processing overhead in hardware/software systems and can
incur increased energy consumption. For example, Bosilca \emph{et.
al.} \cite{bosilca2009algorithm}, report that checkpointing the system
state to detect a single fault per process (and rolling back to a
previous state when faults are detected) leads to $9\%$--$34\%$
time overhead for an implementation using up to $484$ processes.
Similarly, Chen and Dongarra \cite{chen2005fault} report that, for
detecting a single fault per subblock of a large matrix operation,
$4\%$--$9\%$ execution time overhead is incurred in a ScaLAPACK
implementation over a distributed computing system. Finally, Wunderlich
\emph{et. al.} \cite{Wunderlich2013} report that ABFT for the generic
matrix multiply (GEMM) routine incurs $18\%$--$45\%$ execution time
overhead versus the unprotected GEMM on medium to large matrix dimensions
under a GPU implementation. In conjunction with recent studies on
soft errors in processors that indicate that hardware faults tend
to happen in bursts \cite{quinn2011ccc,nicolaidis2012design,alameldeen2011energy},
this shows that ABFT techniques may ultimately not be the best way
to mitigate \emph{arbitrary fault patterns} occurring in 32-bit or
64-bit data representations in memory, arithmetic or logic units of
the utilized hardware. On the other hand, while MR approaches can
indeed mitigate such faults with very high probability, it is well
known that they incur a two-fold or three-fold penalty in execution
time (or energy consumption) as well as substantial data transfers
and latencies to synchronize and cross-check results \cite{engelmann2009case}.

\subsection{Contribution }

We propose a new method to mitigate transient faults or failures in
LSB operations performed in integer data streams with integer arithmetic
units. Examples of such operations are element-by-element additions
and multiplications, inner and outer vector products, sum-of-squares
and permutation operations. They are the building blocks of algorithms
of foundational importance, such as: matrix multiplication \cite{goto2008anatomy,huang1984algorithm},
convolution/cross-correlation \cite{anam2012throughput}, template
matching for search and motion estimation algorithms \cite{kadyrov2006invaders,anastasia2010software,anastasia2012throughput,barbarien2004scalable,foo2008analytical,kontorinis2009statistical,andreopoulos2007adaptive},
covariance calculations \cite{golub1996matrix,yang2004two}, integer-to-integer
transforms \cite{lin2000packed,andreopoulos2002new,andreopoulos2001local,andreopoulos2003high}
and permutation-based encoding systems \cite{fenwick1996burrows},
which form the core of the applications discussed earlier. Our method:
\begin{enumerate}
\item does not generate additional data in form of checksum or duplicate
inputs, as done by ABFT-based or MR-based methods; instead, it performs
pairwise linear superpositions within the numerical representation
of the original inputs, thereby increasing their dynamic range, albeit
in a controllable manner. 
\item does not require modifications to the arithmetic or memory units,
as done by component-based ECC approaches, and can be deployed in
standard 32/64-bit integer units or even 32/64-bit floating-point
units; 
\item does not depend on the specifics of the LSB operation performed; in
fact, it can also be used to detect silent data corruptions in storage
systems, i.e., when no computation is performed with the data. 
\end{enumerate}
Beyond the analytic presentation of our proposal and the theoretical
estimation of its complexity against ABFT, we also present performance
results using fast Fourier transform (FFT) computation, cross-correlation
and matrix product operations, thereby significantly advancing our
early exposition \cite{AnamIOLTS15} that summarized our initial findings
on fail-stop failure mitigation for limited types of operations. The
results show that, for the vast majority of cases, our method's percentile
overhead in execution against the fault-intolerant (i.e., conventional)
realization is upper-bounded by 0.6\%. This overhead is found to be
5 to 1000 times smaller than the one incurred by the ABFT approach
that offers the same fault tolerance capability.

\subsection{Paper Organization }

In \prettyref{sec:ABFT_MR_vs_Entanglement}, we outline ABFT, MR and
the basic concept of the proposed approach for fault tolerance in
numerical stream processing. In \prettyref{sec:From-Numerical-Packing-to-Entanglement}
we present the details for the newly-proposed concept of numerical
entanglement and demonstrate its inherent reliability for LSB processing
of integer streams. \prettyref{sec:Linear_processing} presents the
complexity of numerical entanglements within integer linear or sesquilinear
operations. Section \ref{sec:Experiments} presents experimental comparisons
and \prettyref{sec:Conclusions} presents some concluding remarks.

\section{ABFT/MR Methods versus Numerical Entanglement \label{sec:ABFT_MR_vs_Entanglement}}

\begin{figure*}[tp]
\begin{centering}
\subfigure[Conventional processing]{\includegraphics[scale=0.13]{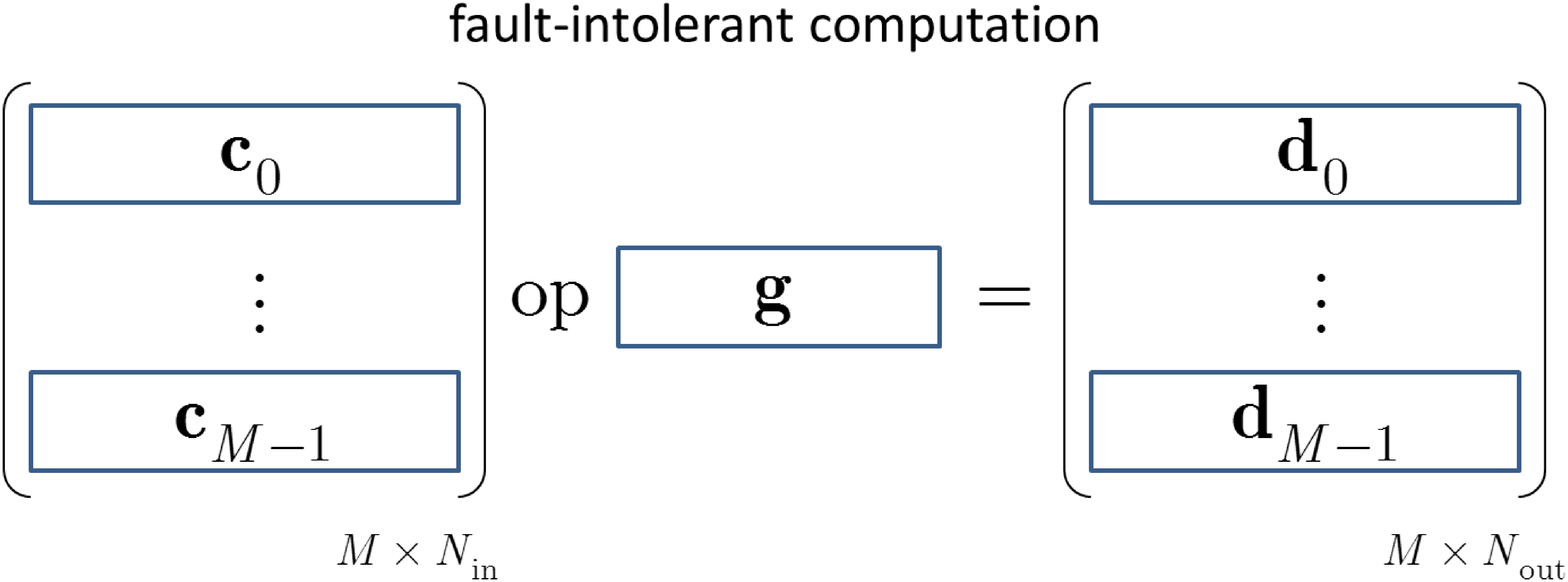}}
\subfigure[ABFT-based processing using one checksum stream]{ \includegraphics[scale=0.13]{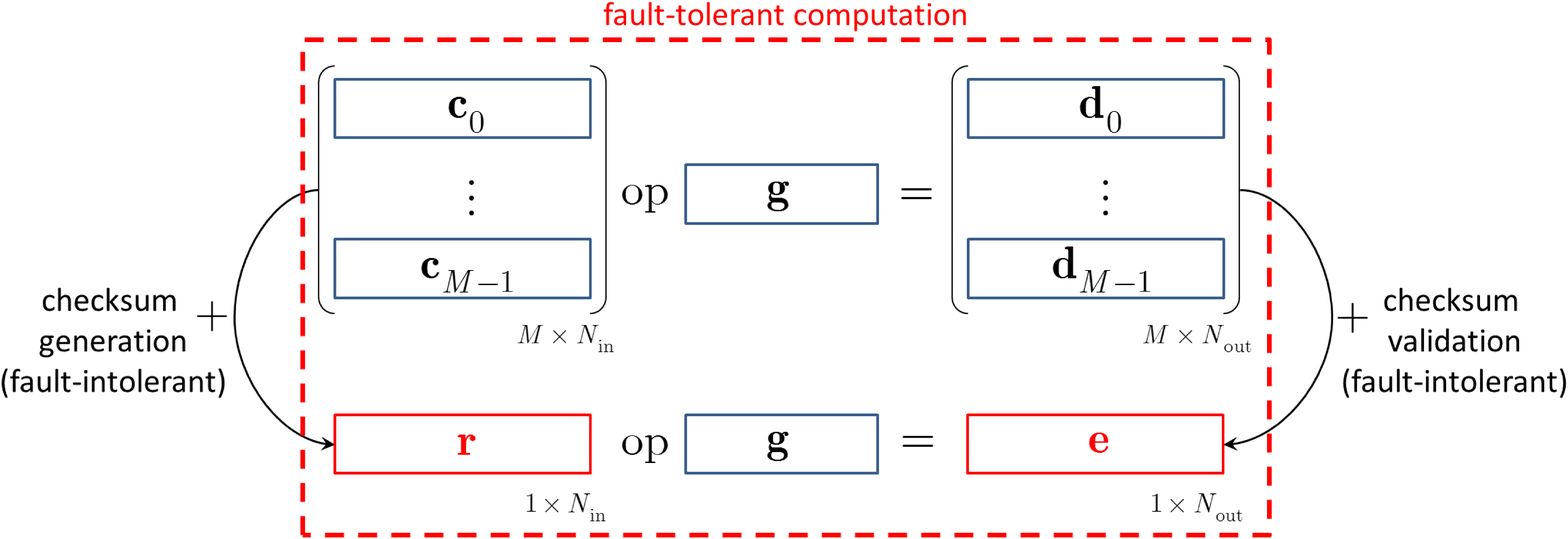}}
\par\end{centering}

\protect\caption{(a) kernel $\mathbf{g}$ applied to $M$ streams of input integers
via LSB operator $\text{op}$; (b) corresponding application using
one checksum input stream for transient fault detection or fail-stop
failure recovery via ABFT. \label{fig:linear_operations_M_input_streams}}
\end{figure*}

Consider a series of $M$ input streams of integers, each comprising
$N_{\text{in}}$ samples%
\footnote{Notations: Boldface uppercase and lowercase letters indicate matrices
and vectors, respectively; the corresponding italicized lowercase
indicate their individual elements, e.g. $\mathbf{A}$ and $a_{m,n}$;
calligraphic uppercase letters indicate operators; $\mathbb{N}{}^{\star}$
is the set of natural numbers excluding zero; $\hat{d}$ denotes the
recovered value of $d$ after unpacking or disentanglement; all indices
are integers. Basic operators: $\left\lfloor a\right\rfloor $ is
the largest integer that is smaller or equal to $a$ (floor operation);
$\left\lceil a\right\rceil $ is the smallest integer that is larger
or equal to $a$ (ceiling operation); $\left\Vert \mathbf{a}\right\Vert $
is vector norm-2; $a\:\&\: b$ denotes binary AND operation between
the bits of $a$ and $b$, respectively; $a\ll b$ and $a\gg b$ indicate
left and right arithmetic shift of integer $a$ by $b$ bits with
truncation occurring at the most-significant or least significant
bit, respectively; $a\,\text{mod}\, b=a-\left\lfloor \frac{a}{b}\right\rfloor b$
is the modulo operation; $a\leftarrow b$ assigns the value of variable
or expression $b$ to variable $a$. %
} ($M\geq3,\: N_{\text{in}}\in\mathfrak{\mathbb{N}{}^{\star}}$): 

\begin{equation}
\mathbf{c}_{m}=\begin{bmatrix}c_{m,0} & \ldots & c_{m,N_{\text{in}}-1}\end{bmatrix},\:0\leq m<M.\label{eq:M_input_data_streams}
\end{equation}
These may be the elements of $M$ rows of a matrix of integers, or
a set of $M$ input integer streams of data to be operated upon with
an integer kernel $\mathbf{g}$. This operation is performed by:

\[
\forall m:\;\mathbf{d}_{m}=\mathbf{c}_{m}\:\text{op}\:\mathbf{g}
\]
\begin{equation}
\mathbf{\text{op}\in\left\{ +\mathrm{,}-\mathrm{,}\times\mathrm{,}\left\langle \centerdot,\centerdot\right\rangle ,\otimes\mathrm{,}\left(\begin{array}{c}
\mathfrak{I}\\
\mathfrak{G}
\end{array}\right)\mathrm{,}\star\right\} }\label{eq:operation_on_inputs}
\end{equation}
with $\mathbf{d}_{m}$ the $m$th vector of output results and $\text{op}$
any LSB operator such as element-by-element addition/subtraction/multiplication,
inner/outer product, permutation%
\footnote{We remark that we consider LSB operations that are \emph{not} data-dependent,
e.g., permutations according to fixed index sets as in the Burrows-Wheeler
transform \cite{fenwick1996burrows,ko2007parameterized}. %
} (i.e., bijective mapping from the sequential index set $\mathbf{\mathfrak{I}}$
to index set $\mathbf{\mathfrak{G}}$ corresponding to $\mathbf{g}$)
and circular convolution or cross-correlation with $\mathbf{g}$.
An illustration of the application of \eqref{eq:operation_on_inputs}
is given in Fig. \ref{fig:linear_operations_M_input_streams}(a).
Beyond the single LSB operator indicated in \eqref{eq:operation_on_inputs}
and illustrated in Fig. \ref{fig:linear_operations_M_input_streams}(a),
we can also assume\emph{ series} of such operators applied consecutively
in order to realize higher-level algorithmic processing, e.g., multiple
consecutive additions, subtractions and scaling operations with pre-established
kernels followed by circular convolutions and permutation operations.
Conversely, the input data streams can also be left in their native
state (i.e., stored in memory), if $\text{op}=\left\{ \times\right\} $
and $\mathbf{g}=1$.

\subsection{Algorithm-based Fault Tolerance and Modular Redundancy}

In their original (or ``pure'') form, the input data streams of
\eqref{eq:M_input_data_streams} are uncorrelated and one input element
cannot be used to cross-check for faults in another without inserting
some form of coding or redundancy. This is conventionally achieved
via ABFT methods \cite{huang1984algorithm,chen2005fault,luk1985weighted,stefanidis2004weighted,nair1988linearCode,sloan2012algorithmic,rexford1992partitioned}.
Specifically, one \emph{additional} input stream is created that comprises
\emph{checksums} of the original inputs:

\begin{equation}
\mathbf{r}=\begin{bmatrix}r_{0} & \ldots & r_{N_{\text{in}}-1}\end{bmatrix},\label{eq:P_redundant_data_streams}
\end{equation}
by using, for example, the sum of the input samples \cite{rexford1992partitioned,sloan2012algorithmic}
at the $n$th position in each of the $M$ streams, $0\leq n<N_{\text{in}}$:

\begin{equation}
\forall n:\; r_{n}=\sum_{m=0}^{M-1}c_{m,n}.\label{eq:redundant_element_definition}
\end{equation}
Then the processing is performed in all input streams $\mathbf{c}_{0},\ldots,\mathbf{c}_{M-1}$
and in the checksum input stream $\mathbf{r}$ by:

\begin{equation}
\begin{bmatrix}\mathbf{d}_{0}\\
\vdots\\
\mathbf{d}_{M-1}\\
\mathbf{e}
\end{bmatrix}=\begin{bmatrix}\mathbf{c}_{0}\\
\vdots\\
\mathbf{c}_{M-1}\\
\mathbf{r}
\end{bmatrix}\:\text{op}\:\mathbf{g},\label{eq:operation_on_inputs_and_redundant_inputs}
\end{equation}
Any transient faults in any single stream out of $M+1$ output streams
can then be detected by checking if:

\begin{equation}
\exists n:\;\sum_{m=0}^{M-1}d_{m,n}\neq e_{n}.\label{eq:error_checking}
\end{equation}
This process is pictorially illustrated in Fig. \ref{fig:linear_operations_M_input_streams}(b).
Similarly, result recovery after any single fail-stop failure can
take place by subtracting the results of all the remaining output
streams from $\mathbf{r}$. As discussed in partitioning schemes for
checksum-based methods and ABFT \cite{rexford1992partitioned,sloan2012algorithmic},
the recovery capability can be increased by using additional weighted
checksums. However, this comes at the cost of increasing the number
of checksum input streams, which leads to increased overhead. For
this reason, practical approaches tend to use a single checksum stream
\cite{bosilca2009algorithm,chen2005fault,murray2008spread,huang1984algorithm,sloan2012algorithmic}.
At the other extreme, when $M$ checksum streams are used, this corresponds
to repeating the operation twice (dual modular redundancy) and any
fault on the original computation can be detected if the results are
compared with the results of the checksum set. In summary, the practical
limitations of ABFT are: 
\begin{enumerate}
\item The percentile implementation overhead (i.e., processing cycles, energy
consumption, memory accesses) of ABFT is $\frac{1}{M}\times100\%$. 
\item The dynamic range of the computations with each of the checksum input
streams is increased by $\left\lceil \log_{2}M\right\rceil $ bits,
as each of the checksum input data values comprises the sum of groups
of $M$ input samples, as shown in \eqref{eq:redundant_element_definition}. 
\item The overall execution flow changes as the total number of processed
streams is changed from $M$ to $M+1$. 
\end{enumerate}

\subsection{Numerical Entanglement}

\begin{figure}[tp]
\begin{centering}
\includegraphics[scale=0.13]{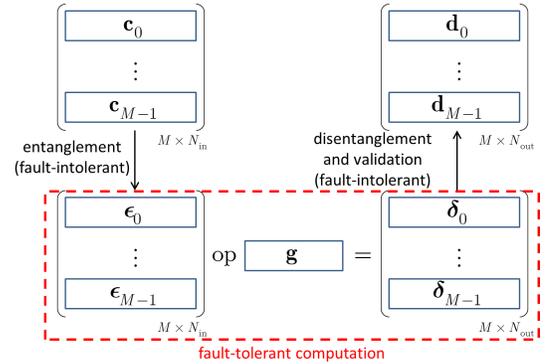}
\par\end{centering}

\protect\caption{LSB processing via numerical entanglement, followed by result recovery
and validation. \label{fig:numerical_entanglement_overview}}
\end{figure}

In our proposal, numerical entanglement mixes the inputs prior to
linear processing using linear superposition and ensures the results
can be extracted and validated via a mixture of shift-add operations.
It is conceptualized in Fig. \ref{fig:numerical_entanglement_overview}.
As shown there, $M$ ($M\geq3$) input streams (each comprising $N_{\text{in}}$
integer samples and denoted by $\mathbf{c}_{m}$ $0\leq m<M$) become
$M$ entangled streams of integers (of $N_{\text{in}}$ integer samples
each), $\boldsymbol{\epsilon}_{m}$. Each element of the $m$th entangled
stream, $\epsilon_{m,n}$ ($0\leq n<N_{\text{in}}$), comprises the
superposition of two input elements $c_{x,n}$ and $c_{y,n}$ from
different input streams $x$ and $y$, i.e., $0\leq x,y<M$ and $x\neq y.$
The LSB operation is carried out with the entangled streams, thereby
producing the entangled output streams $\boldsymbol{\delta}_{m}$
(each comprising $N_{\text{out}}$ integer samples). These can be
disentangled to recover the final results $\mathbf{\hat{d}}_{m}$.
Any transient faults that occurred on any single entangled output
stream out of $M$ are detectable with a single test that utilizes
additions and shift operations. In addition, any single fail-stop
failure can be mitigated from the results of the remaining streams. 

Unlike checksum or MR methods, numerical entanglement does not use
additional checksum streams. Therefore, the complexity of entanglement,
disentanglement (recovery) and fault checking does not depend on the
complexity of the operator $\text{op}$, or on the length and type
of the kernel (operand) $\mathbf{g}$. The entangled inputs can be
written in-place and no additional storage or additional operations
are needed during the execution of the actual operation (albeit at
the cost of reducing the dynamic range supported). In fact, with the
exception of addition/subtraction operations with constants, no modifications
are performed to the processors, software routines and arithmetic
units performing the operation with kernel $\mathbf{g}$ and the computational
system design remains unaware of the fact that entangled input streams
are used instead of the original input streams. Thus, the entangled
computation shown in Fig. \ref{fig:numerical_entanglement_overview}
can be executed concurrently in $M$ processing cores (that may be
physically separate) and any memory optimization or other algorithmic
optimization can be applied in the same manner as for the original
computation. For example, if a fast Fourier transform (FFT) routine
is used for the calculation of convolution or cross-correlation of
each input stream $\mathbf{c}_{m}$ with kernel $\mathbf{g}$, this
routine can be used directly with the entangled input streams $\boldsymbol{\epsilon}_{m}$
and kernel $\mathbf{g}$. A summary of the features of each approach
is presented in Table \ref{tab:Summary-of-features}. The comparison
includes our previous work on numerical packing and duplicate execution
\cite{anastasia2010linear,anastasia2012throughput,anam2012throughput,AnaradoIOLTS,AnaradoIOLTS15},
which can be seen as an alternative form of MR. 

\begin{table*}[tbh]
\protect\caption{Summary of features of different methods for fault tolerance within
each group of $M$ outputs of $M$ streams, with each stream element
comprising $w$ bits. \label{tab:Summary-of-features}}

\centering{}%
\begin{tabular}{|c|c|c|c|c|}
\hline 
Method & ABFT  & Dual Modular  & Packing and Duplicate  & Proposed \tabularnewline
\cline{1-1} 
Feature & \cite{bosilca2009algorithm,chen2005fault,huang1984algorithm,murray2008spread} & Redundancy \cite{engelmann2009case} & Execution \cite{anastasia2010linear,anastasia2012throughput,anam2012throughput} & Numerical Entanglement\tabularnewline
\hline 
\hline 
\noalign{\vskip0.03cm}
In-place storage & No & No & No & Yes\tabularnewline
\hline 
\% of redundant & \multirow{2}{*}{$\frac{1}{M}\times100\%$ } & \multirow{2}{*}{100\%} & \multirow{2}{*}{0\%} & \multirow{2}{*}{0\%}\tabularnewline
computations &  &  &  & \tabularnewline
\hline 
Output bitwidth reduction due  & $\left\lceil \log_{2}M\right\rceil $ bits (only  & \multirow{2}{*}{0 bits} & \multirow{2}{*}{0 bits} & \multirow{2}{*}{$\left\lceil \frac{w}{M}\right\rceil $ bits}\tabularnewline
to fault tolerance capability & for checksum inputs)  &  &  & \tabularnewline
\hline 
Guaranteed fault detection / & 1 fault in every  & 1 fault in every & 1 fault in every & 1 fault in every\tabularnewline
fail-stop failure mitigation &  $M$ outputs  & 2 outputs & 2 outputs & $M$ outputs\tabularnewline
\hline 
Percentile overhead & \multirow{2}{*}{$\frac{1}{M}\times100\%$} & More than & More than & $0.03\%$ to $7\%$ (decreases\tabularnewline
in execution &  & $100\%$ & $100\%$ & with operand length)\tabularnewline
\hline 
\end{tabular}
\end{table*}

\section{Numerical Entanglement \label{sec:From-Numerical-Packing-to-Entanglement}}

Numerical entanglement bears some resemblance to the concept of numerical
packing proposed previously by Andreopoulos \emph{et. al.} \cite{anastasia2010linear,anastasia2012throughput,anam2012throughput,anastasia2010software,AnaradoIOLTS,AnaradoIOLTS15},
Kadyrov and Petrou \cite{kadyrov2006invaders} and others \cite{lin2000packed,allen1996approach}.
By using multiple packed representations and no overlap, it can be
shown \cite{AnaradoIOLTS15} that all faults occurring on a single
description can be detected, at the cost of using a 64-bit integer
representation that can accommodate up to 19-bit signed integer outputs
in packed format. This illustrates that utilizing packing for fault
detection may be a viable approach, but it comes at the cost of significant
reduction in the dynamic range supported by the packed representation.
The proposed approach overcomes this limitation by allowing inputs
to superimpose each other, thereby creating an input representation
where certain bits from pairs of input data samples are numerically
entangled, as explained in the following.

\subsection{Proposed Numerical Entanglement in Groups of Three Inputs ($M=3$)}

Numerical entanglement guarantees the detection of \emph{any transient
fault} occurring in any one out of the $M$ entangled streams created.
In addition, if a separate core is used for the processing of each
stream, it allows for the recovery of all outputs after any single
fail-stop core failure.

\subsubsection{Entanglement}

\begin{figure*}[tp]
\begin{centering}
\includegraphics[scale=0.22]{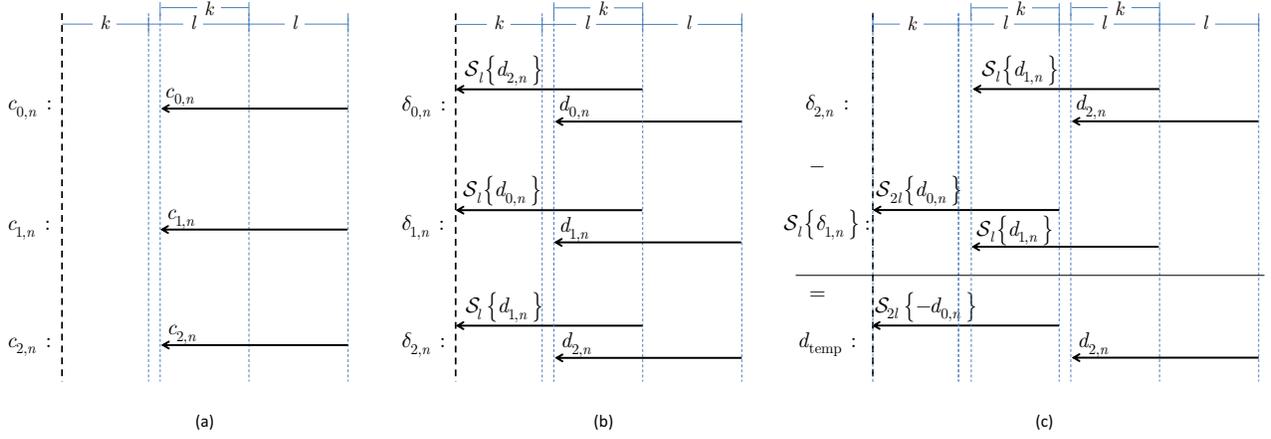}
\par\end{centering}

\protect\caption{The arrows indicate the dynamic range of the represented data elements
(from least-significant to most-significant bits). (a) Illustration
of three input data samples at position $n$; (b) entangled outputs
after entangling and integer LSB processing; (c) illustration of the
first part of \eqref{eq:integer_disentanglement} that produces the
temporary variable $d_{\text{temp}}$. \label{fig:Entanglement-with-M_equal_3} }
\end{figure*}

In the simplest form of entanglement ($M=3$), each triplet of input
samples {[}shown in Fig. \ref{fig:Entanglement-with-M_equal_3}(a){]}
of the three integer streams, $c_{0,n}$, $c_{1,n}$ and $c_{2,n}$,
$0\leq n<N_{\text{in}}$, produces the following entangled triplet
via the linear superposition:

\begin{eqnarray}
\mathit{\epsilon}_{0,n} & = & \mathcal{S}_{l}\left\{ c_{2,n}\right\} +c_{0,n}\nonumber \\
\epsilon_{1,n} & = & \mathcal{S}_{l}\left\{ c_{0,n}\right\} +c_{1,n}\label{eq:entanglement_M_equal_1}\\
\epsilon_{2,n} & = & \mathcal{S}_{l}\left\{ c_{1,n}\right\} +c_{2,n}\nonumber 
\end{eqnarray}
with 

\begin{equation}
2l+k\leq w\label{eq:entanglement-condition-M_eq_1}
\end{equation}
where $w\in\left\{ 32,64\right\} $ for 32 or 64-bit integer representations,
and

\begin{equation}
\mathcal{S}_{b}\left\{ a\right\} \equiv\left\{ \begin{array}{c}
\;\;\left(a\ll b\right),\;\;\;\;\,\text{if}\; b\geq0\\
\left[a\gg\left(-b\right)\right],\;\text{if}\; b<0
\end{array}\right.
\end{equation}
the left or right arithmetic shift of $a$ by $b$ bits, with the
maximum dynamic range supported for each signed input, $c_{0,n}$,
$c_{1,n}$ and $c_{2,n}$, being proportional to $l+k$ bits.

The values for $l$ and $k$ are chosen such that $l+k$ is maximum
within the constraint of \eqref{eq:entanglement-condition-M_eq_1}
and $k\leq l$. Via the application of LSB operations, each $\boldsymbol{\epsilon}_{m}$
entangled input stream ($0\leq m<M$) is converted to the entangled
output stream%
\footnote{For the particular cases of: $\text{op}\in\left\{ +,-\right\} $,
$\mathbf{g}$ must also be entangled with itself via: $g_{n}\leftarrow\mathcal{S}_{l}\left\{ g_{n}\right\} +g_{n}$,
in order to retain the homomorphism of the performed operation. All
other operations occur without any modification in $\mathbf{g}$.%
} $\boldsymbol{\delta}_{m}$ (which contains $N_{\text{out}}$ values): 

\begin{equation}
\forall m:\boldsymbol{\;\delta}_{m}=\left(\boldsymbol{\epsilon}_{m}\:\text{op}\:\mathbf{g}\right).\label{eq:operation_on_inputs-1-1}
\end{equation}
A conceptual illustration of the entangled outputs after \eqref{eq:entanglement_M_equal_1}
and \eqref{eq:operation_on_inputs-1-1} is given in Fig. \ref{fig:Entanglement-with-M_equal_3}(b).
The overlap of the arrows in the representations of $\delta_{0,n}$,
$\delta_{1,n}$ and $\delta_{2,n}$ indicate the ``entangled'' region
of $k$ bits, where the final outputs, $d_{0,n}$, $d_{1,n}$ and
$d_{2,n}$, numerically superimpose each other (in pairs) due to the
shift and add operations of \eqref{eq:entanglement_M_equal_1}. Similarly
as before, $l$ bits of dynamic range are sacrificed in order to detect
faults (or mitigate a fail-stop failure) during the computation of
\eqref{eq:operation_on_inputs-1-1}. Since it is assumed that the
dynamic range of the inputs does not exceed $l+k$ bits, the entangled
representation is contained within $2l+k$ bits and never overflows.
As a practical instantiation of \eqref{eq:entanglement_M_equal_1},
we can set $w=32$, $l=11$ and $k=10$ in a signed 32-bit integer
configuration. 

We now describe the disentanglement and result validation (or recovery)
process. The reader can also consult Fig. \ref{fig:Entanglement-with-M_equal_3}.

\subsubsection{Disentanglement}

We can disentangle and recover the final results $\hat{d}_{0,n}$,
$\hat{d}_{1,n}$ and $\hat{d}_{2,n}$ by ($0\leq n<N_{\text{out}}$): 

\begin{eqnarray}
d_{\textrm{temp}} & = & \delta_{2,n}-\mathcal{S}_{l}\left\{ \delta_{1,n}\right\} \nonumber \\
\hat{d}_{2,n} & = & \mathcal{S}_{-2\left(w-l\right)}\left\{ \mathcal{S}_{2\left(w-l\right)}\left\{ d_{\textrm{temp}}\right\} \right\} \nonumber \\
\hat{d}_{0,n} & = & \mathcal{S}_{-2l}\left\{ -\left(d_{\textrm{temp}}-\hat{d}_{2,n}\right)\right\} \label{eq:integer_disentanglement}\\
\hat{d}_{1,n} & = & \delta_{1,n}-\mathcal{S}_{l}\left\{ \hat{d}_{0,n}\right\} \nonumber 
\end{eqnarray}
The first three parts of \eqref{eq:integer_disentanglement} assume
a $2w$-bit integer representation is used for the interim operations,
as the temporary variable $d_{\textrm{temp}}$ is stored in $2w$-bit
integer representation. However, all recovered results, $\hat{d}_{0,n}$,
$\hat{d}_{1,n}$ and $\hat{d}_{2,n}$, require only $l+k$ bits. 

Explanation of \eqref{eq:integer_disentanglement}: The first part
creates a temporary composite number, $d_{\text{temp}}$, comprising
$\hat{d}_{0,n}$ in the $l+k$ most-significant bits and $\hat{d}_{2,n}$
in the $2l$ least-significant bits (therefore, $d_{\textrm{temp}}$
requires $3l+k$ bits). The creation of $d_{\text{temp}}$ is pictorially
illustrated in Fig. \ref{fig:Entanglement-with-M_equal_3}(c). In
the second part, $\hat{d}_{2,n}$ is extracted by: \emph{(i)} discarding
the $\left(2w-2l\right)$ most-significant bits of $d_{\text{temp}}$;
\emph{(ii)} arithmetically shifting the output down to the correct
range. The third part of \eqref{eq:integer_disentanglement} uses
$\hat{d}_{2,n}$ to recover $\hat{d}_{0,n}$ from $d_{\text{temp}}$
and, in the fourth part of \eqref{eq:integer_disentanglement}, $\hat{d}_{0,n}$
is used to recover $\hat{d}_{1,n}$. Notice that \eqref{eq:integer_disentanglement}
recovers all three $\hat{d}_{0,n}$, $\hat{d}_{1,n}$ and $\hat{d}_{2,n}$
without using $\delta_{0,n}$. This is a crucial aspect that leads
to the fault tolerance characteristic of our proposal, which are discussed
next.

\subsection{Properties and Fault Tolerance Characteristics}

\emph{Remark 1 (operations within $w$ bits):} To facilitate our exposition,
the first three parts of \eqref{eq:integer_disentanglement} are presented
under the assumption of a $2w$-bit integer representation. However,
it is straightforward to implement them via $w$-bit integer operations
by separating $d_{\text{temp}}$ into two parts of $w$ bits and performing
the operations separately within these parts\emph{.}

\emph{Remark 2 (dynamic range):} Bit $l+k$ within each recovered
output $\hat{d}_{0,n}$, $\hat{d}_{1,n}$ and $\hat{d}_{2,n}$ represents
its sign bit. Given that: \emph{(i)} each entangled output comprises
the addition of two outputs (with one of them left-shifted by $l$
bits); \emph{(ii)} the entangled outputs must not exceed $2l+k$ bits,
the outputs of the LSB operations must not exceed the range 

\begin{equation}
\forall n:\; d_{0,n},d_{1,n},d_{2,n}\in\left\{ -\left(2^{l+k-1}-2^{l}\right),\ldots,2^{l+k-1}-2^{l}\right\} .\label{eq:dynamic_range}
\end{equation}
Therefore, \eqref{eq:dynamic_range} comprises the range permissible
for the LSB operations with the entangled representation. 
\begin{prop}
If a separate core is used for each stream computation of \eqref{eq:operation_on_inputs-1-1}
with $M=3$, the disentanglement process of \eqref{eq:integer_disentanglement}
can recover all results, $\mathbf{\hat{d}}_{0}$, $\mathbf{\hat{d}}_{1}$,
and $\mathbf{\hat{d}}_{2}$, after any single fail-stop failure.\end{prop}
\begin{IEEEproof}
See Appendix.
\end{IEEEproof}
The following proposition proves that, if all three entangled output
streams are available, we can detect any transient fault in any single
stream. 
\begin{prop}
Any transient fault occurring on a single entangled output stream
during the computation of \eqref{eq:operation_on_inputs-1-1} with
$M=3$ is detectable. \end{prop}
\begin{IEEEproof}
See Appendix.
\end{IEEEproof}

\subsection{Generalized Entanglement in Groups of $M$ Inputs ($M\geq3$)}

We extend the proposed entanglement process to using $M$ inputs and
providing $M$ entangled descriptions, each comprising the linear
superposition of two inputs. This ensures that, for every\emph{ }$n$
($0\leq n<N_{\text{out}}$)\emph{, any }single transient fault will
be detected within each group of $M$ output samples. Alternatively,
if $M$ separate cores are used for the computation of the $M$ entangled
outputs, any single fail-stop failure will be mitigated from the remaining
$M-1$ output entangled streams. 

The condition for ensuring that overflow is avoided is

\begin{equation}
\left(M-1\right)l+k\leq w\label{eq:M_l_k_constraint}
\end{equation}
and the dynamic range supported for all outputs is ($\forall m,n$): 

\begin{equation}
d_{m,n}\in\left\{ -2^{\left(M-3\right)l+k}\left(2^{l-1}-1\right),\ldots,2^{\left(M-3\right)l+k}\left(2^{l-1}-1\right)\right\} .\label{eq:M_dynamic_range}
\end{equation}
The values for $l$ and $k$ are chosen such that $\left(M-2\right)l+k$
is maximum within the constraint of \eqref{eq:M_l_k_constraint} and
$k\leq l$. 

We now define the following circulant matrix operator comprising cyclic
permutations of the $1\times M$ vector $\begin{bmatrix}1 & 0 & \cdots & 0 & \mathcal{S}_{l}\end{bmatrix}$:

\begin{equation}
\boldsymbol{\mathcal{E}}=\begin{bmatrix}1 & 0 & \cdots & 0 & \mathcal{S}_{l}\\
\mathcal{S}_{l} & 1 & \cdots & 0 & 0\\
 &  & \ddots\\
0 & \cdots & \mathcal{S}_{l} & 1 & 0\\
0 & \cdots & 0 & \mathcal{S}_{l} & 1
\end{bmatrix}_{M\times M}.\label{eq:entangle_matrix}
\end{equation}
Operator $\boldsymbol{\mathcal{E}}$ generalizes the proposed numerical
entanglement process. Specifically, following the case of $M=3$,
in the generalized entanglement in groups of $M$ streams, two inputs
are entangled together (with one of the two shifted by $l$ bits)
to create each entangled input stream of data. Any LSB operation is
then performed on these entangled input streams and\emph{ }we shall
show that any\emph{ }transient fault occurring during the processing
of a single entangled stream can be detected within each group of
$M$ outputs. Alternatively, any single fail-stop core failure can
be mitigated. 

\begin{figure}[tp]
\begin{centering}
\includegraphics[scale=0.36]{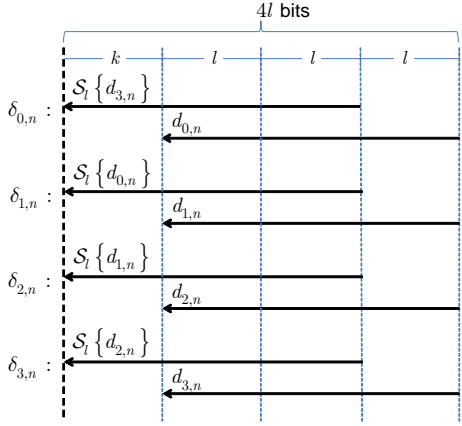}
\par\end{centering}

\protect\caption{Illustration of entanglement of $M=4$ outputs after integer LSB processing.
In this case, for $w\in\left\{ 32,64\right\} $, $k=l=\frac{w}{4}$.
\label{fig:Entanglement-with-M_equal_4} }
\end{figure}

\begin{figure}[tbh]
\begin{centering}
\includegraphics[scale=0.36]{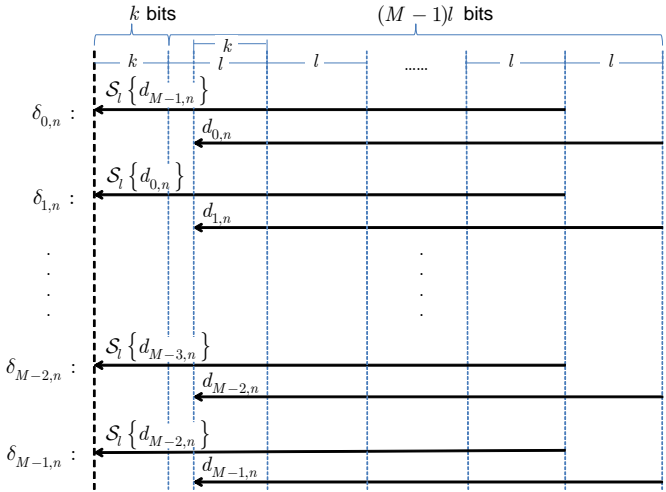}
\par\end{centering}

\protect\caption{Illustration of the general form of entanglement of $M$ outputs after
integer LSB processing. \label{fig:M_part1}}
\end{figure}

For every input stream position $n$, $0\leq n<N_{\text{in}}$, the
entanglement vector performing the linear superposition of pairs out
of $M$ inputs is now formed by:

\begin{equation}
\begin{bmatrix}\epsilon_{0,n} & \cdots & \epsilon_{M-1,n}\end{bmatrix}^{\text{T}}=\mathcal{\boldsymbol{\mathcal{E}}}\left\{ \begin{bmatrix}c_{0,n} & \cdots & c_{M-1,n}\end{bmatrix}^{\text{T}}\right\} \label{eq:entanglement_M_general}
\end{equation}
Fig. \ref{fig:Entanglement-with-M_equal_4} illustrates the entangled
outputs after \eqref{eq:entanglement_M_general} and the LSB processing
of \eqref{eq:operation_on_inputs-1-1} under $M=4$. In addition,
Fig. \ref{fig:M_part1} illustrates the general case of $M$ outputs,
produced after $M$ inputs were entangled to create $M$ descriptions
and \eqref{eq:entanglement_M_general} and \eqref{eq:operation_on_inputs-1-1}
were performed.

After the application of \eqref{eq:operation_on_inputs-1-1}, we can
disentangle every output stream element $\delta_{m,n}$, $0\leq m<M$,
$0\leq n<N_{\text{out}}$, as follows. Let us first identify the entangled
output stream $\boldsymbol{\delta}_{r}$ (with $0\leq r<M$) that
we shall not use in the disentanglement process, either because it
is unavailable due a fail-stop failure, or because we would like to
recreate it if we are checking for transient faults. We produce the
$2w$-bit temporary variable%
\footnote{Due to the usage of $2w$ bits, \eqref{eq:integer_disentanglement_partial_M_general}--\eqref{eq:d_hat_0,n}
must be separated in two parts of $w$ bits if the entire operation
has to occur via $w$-bit integer arithmetic. Since this is an implementation
issue, we do not illustrate this separation in our exposition. %
} $d_{\textrm{temp}}$ by:

\begin{equation}
d_{\textrm{temp}}=\sum_{m=0}^{M-2}\left(-1\right)^{m}\mathcal{S}_{(M-2-m)l}\left\{ \delta_{\left(r+1+m\right)\text{mod}M,n}\right\} .\label{eq:integer_disentanglement_partial_M_general}
\end{equation}

\begin{figure}[tbh]
\begin{centering}
\includegraphics[scale=0.36]{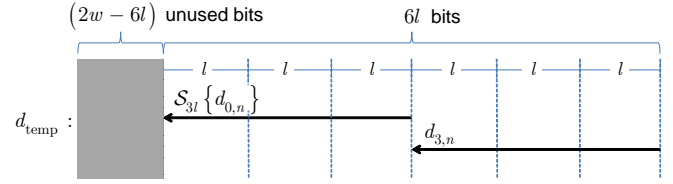}
\par\end{centering}

\protect\caption{Temporary value $d_{\textrm{temp}}$ produced during the disentanglement
process of \eqref{eq:integer_disentanglement_partial_M_general} within
an integer representation, with $M=4$. As shown in Table \ref{tab:Examples_l_k},
for $w\in\left\{ 32,64\right\} $, $k=l$ for this case. \label{fig:Entanglement-with-M_equal_4_2} }
\end{figure}

An illustration of the result of \eqref{eq:integer_disentanglement_partial_M_general}
for $M=4$ and $r=0$ is given in Fig. \ref{fig:Entanglement-with-M_equal_4_2}.
Notice that \eqref{eq:integer_disentanglement_partial_M_general}
does not use $\boldsymbol{\delta}_{r}$. We can then extract the value
of $\hat{d}_{r,n}$ and $\hat{d}_{\left(r+M-1\right)\text{mod}M,n}$
directly from $d_{\textrm{temp}}$:

\begin{equation}
\hat{d}_{\left(r+M-1\right)\text{mod}M,n}=\mathcal{S}_{-\left[2w-\left(M-1\right)l\right]}\left\{ \mathcal{S}_{2w-\left(M-1\right)l}\left\{ d_{\textrm{temp}}\right\} \right\} \label{eq:integer_signed_disentanglement_partial_M_general}
\end{equation}

\begin{equation}
\hat{d}_{r,n}=\mathcal{S}_{-\left(M-1\right)l}\left\{ \left(-1\right)^{M}\left(d_{\textrm{temp}}-\hat{d}_{M-1,n}\right)\right\} .\label{eq:d_hat_0,n}
\end{equation}
The other outputs can now be disentangled by ($1\leq m<M-2$): 

\begin{equation}
\hat{d}_{\left(r+m\right)\text{mod}M,n}=\mathrm{\mathbf{\delta}}_{\left(r+m\right)\text{mod}M,n}-\mathcal{S}_{l}\left\{ \hat{d}_{\left(r+m-1\right)\text{mod}M,n}\right\} .\label{eq:integer_disentanglement_M_general-1}
\end{equation}

\begin{prop}
If a separate core is used for each stream computation of \eqref{eq:operation_on_inputs-1-1},
the disentanglement process of \eqref{eq:integer_disentanglement_partial_M_general}--\eqref{eq:integer_disentanglement_M_general-1}
can recover all results, $\mathbf{\hat{d}}_{0},\ldots,\mathbf{\hat{d}}_{M-1}$,
after any single fail-stop failure.\end{prop}
\begin{IEEEproof}
See Appendix. \end{IEEEproof}
\begin{prop}
Any transient fault occurring on a single entangled stream during
the computation of \eqref{eq:operation_on_inputs-1-1} with $M\geq3$
is detectable. \end{prop}
\begin{IEEEproof}
See Appendix.
\end{IEEEproof}
\emph{Remark 3 (dynamic range of generalized entanglement and equivalence
to checksum methods):} Examples for the maximum bitwidth achievable
for different cases of $M$ are given in Table \ref{tab:Examples_l_k}
assuming a 32-bit representation. We also present the dynamic range
permitted by the equivalent checksum-based method {[}\eqref{eq:P_redundant_data_streams}--\eqref{eq:error_checking}{]}
in order to ensure that its checksum stream does not overflow under
a 32-bit representation. Evidently, for $M\leq10$, the proposed approach
incurs loss of $1$ to $9$ bits of dynamic range against the checksum-based
method, while it allows for higher dynamic range than the checksum-based
method for $M\geq11$. At the same time, our proposal does not require
the overhead of applying the LSB operations to an additional stream,
as it ``overlays'' the information of each input onto another input
via the numerical entanglement of pairs of inputs. Beyond this important
difference, Propositions 3 and 4 show that our approach offers the
exact equivalent to the checksum method of \eqref{eq:P_redundant_data_streams}--\eqref{eq:error_checking}
for integer inputs.

\emph{Remark 4 (extensions):} It is of theoretical interest to consider
whether the proposed numerical entanglement approach can be extended
to guarantee detection of multiple transient faults occurring in co-located
positions in the entangled streams, or recover from more than a single
fail-stop failure in $M$ streams. In addition, while LSB operations
cover a wide range of compute- and memory-intensive DSP systems, it
would be interesting to investigate the applicability of our approach
to data-dependent operations or non-linear operations, like modulo,
binary operators, etc. Finally, beyond integer processing, the fault
tolerance of numerical entanglement would be extremely beneficial
to floating-point LSB operations. 

Concerning the first point, a potential solution could be to create
multiple tiers of entanglement, i.e., reapply the operator of \eqref{eq:entanglement_M_general}
to $\begin{bmatrix}\epsilon_{0,n} & \cdots & \epsilon_{M-1,n}\end{bmatrix}^{\text{T}}$
to create $K$ entanglement stages, and thereby examine if this leads
to the possibility of resilience to $K$ transient faults in $M$
co-located entangled outputs (or recovery from $K$ fail-stop failures
in $M$ streams) without requiring $K$ checksum streams like ABFT.
While we have some encouraging results in this direction, due to space
limitations we plan to investigate and quantify this in another paper. 

The second point (extension beyond LSB operations) is also a limitation
of ABFT fault-tolerance methods. Therefore, it can be addressed in
a similar way as carried out for such cases, i.e., disentangle all
outputs before the non-LSB operation, carry out this operation with
the disentangled outputs (using another form of fault tolerance---such
as MR---for this process), and then re-entangle the results if no
faults or failures are detected. This requires a system-level approach
to carefully leverage the cost of such modifications, so we opt to
leave it as a topic for future work. 

Finally, concerning extension to floating-point arithmetic, we do
not foresee a direct way to achieve this with the proposed approach,
as standard floating-point arithmetic does not allow for two ``clean''
top and bottom zones of bits in the way presented in Figs. \ref{fig:Entanglement-with-M_equal_3},
\ref{fig:Entanglement-with-M_equal_4} and \ref{fig:M_part1}. Moreover,
it is important to note that, in the case of floating-point arithmetic,
even conventional ABFT approaches require more complex handling of
fault detection based on appropriate thresholds, as the sum of the
results of the $M$ output streams will not be reproduced exactly
by the results of the checksum stream due to the lossy and non-commutative
nature of floating-point arithmetic. However, possibilities to incorporate
kernel-adaptive companding and rounding (that convert floating-point
inputs into integers for error-tolerant generic matrix multiplication
kernels \cite{anastasia2012throughput}) prior to entanglement, may
be investigated in future work. 

\begin{table}[tbh]
\protect\caption{Examples of $l$ and $k$ values and bitwidth supported for the output
data under $w=32$ bits and: \emph{(i)} different numbers of entanglements;
\emph{(ii)} ABFT with $P=1$. Both approaches guarantee the mitigation
of a fail-stop failure (or detect transient faults) in one out of
$M$ streams. \label{tab:Examples_l_k}}

\centering{}%
\begin{tabular}{|c|c|c|c|c|}
\hline 
\multirow{3}{*}{$M$} & \multirow{3}{*}{$l$} & \multirow{3}{*}{$k$} & \multicolumn{2}{c|}{Maximum bitwidth supported by}\tabularnewline
 &  &  & proposed: & ABFT:\tabularnewline
 &  &  & $\left(M-2\right)l+k$ & $w-\left\lceil \log_{2}M\right\rceil $\tabularnewline
\hline 
3 & 11 & 10 & 21 & 30\tabularnewline
\hline 
4 & 8 & 8 & 24 & 30\tabularnewline
\hline 
5 & 7 & 4 & 25 & 29\tabularnewline
\hline 
8 & 4 & 4 & 28 & 29\tabularnewline
\hline 
11 & 3 & 2 & 29 & 28\tabularnewline
\hline 
16 & 2 & 2 & 30 & 28\tabularnewline
\hline 
32 & 1 & 1 & 31 & 27\tabularnewline
\hline 
\end{tabular}
\end{table}

\section{Complexity in LSB Operations with Numerical Entanglements\label{sec:Linear_processing}}

We now turn our attention to the cost of performing numerical entanglement,
result extraction and validation versus the cost of the LSB operation
itself. 

\begin{figure*}[tbh]
\begin{centering}
\subfigure[GEMM]{\includegraphics[scale=0.13]{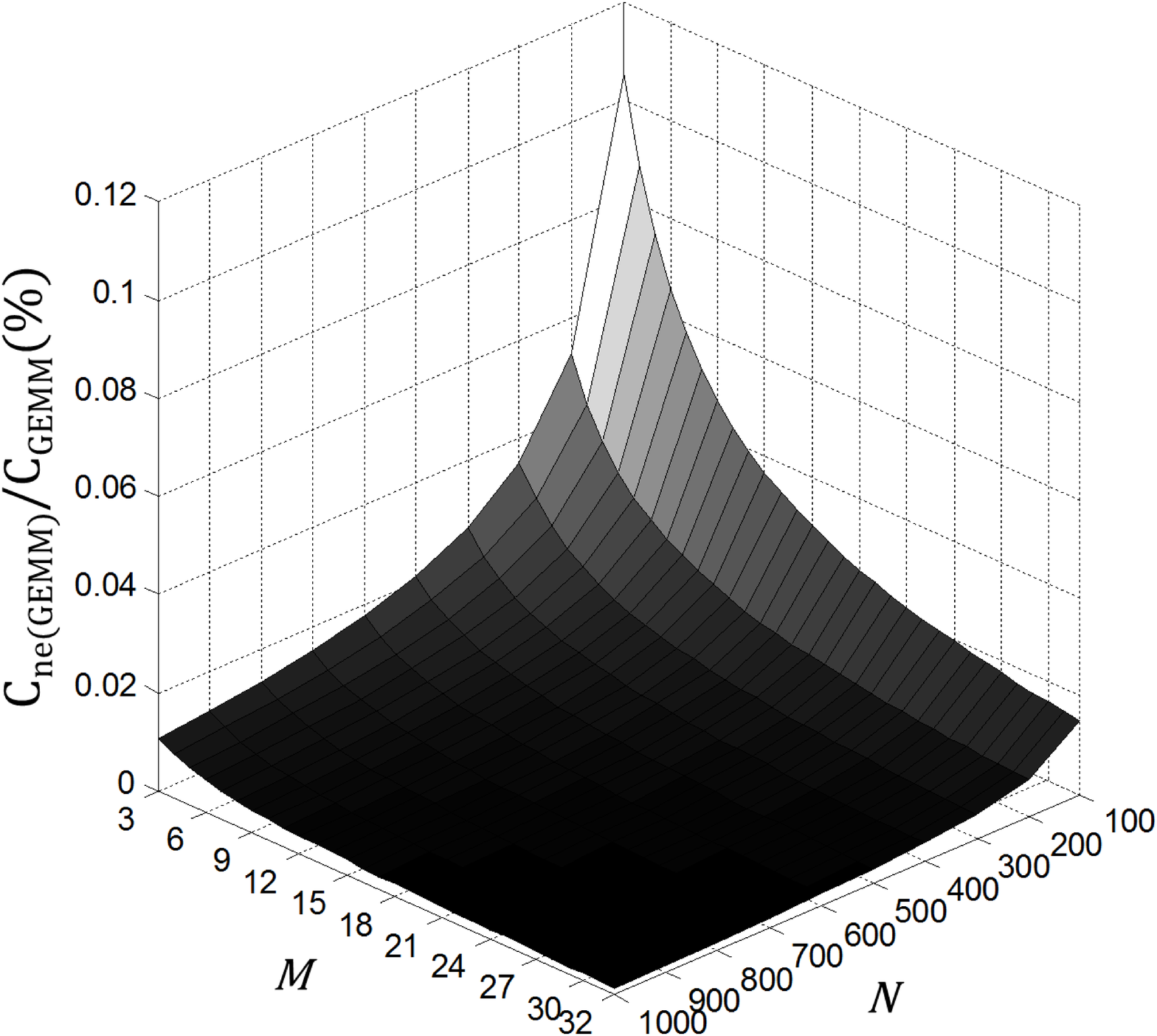}}
\subfigure[Convolution (time)]{\includegraphics[scale=0.13]{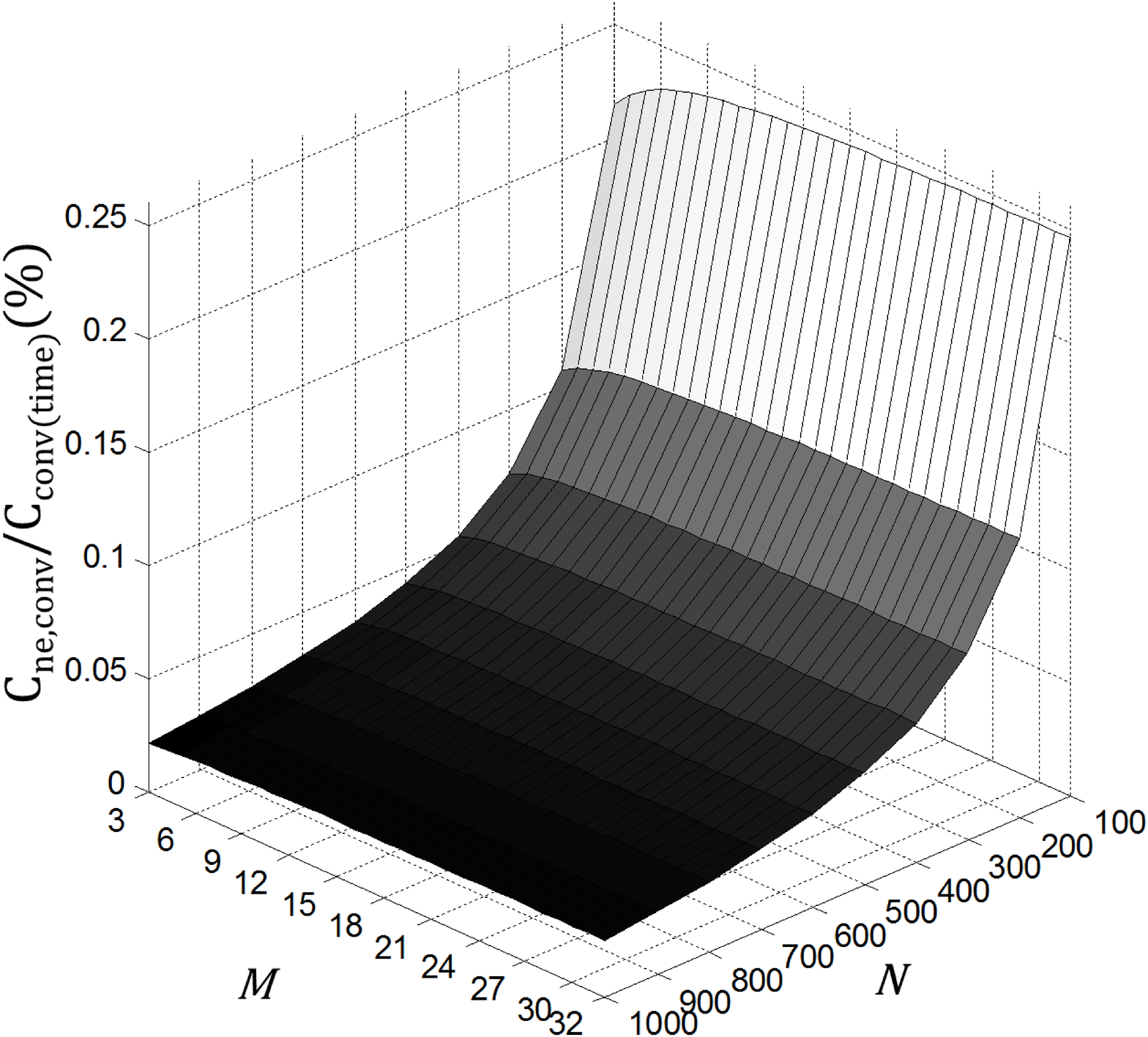}}\subfigure[Convolution (freq)]{\includegraphics[scale=0.13]{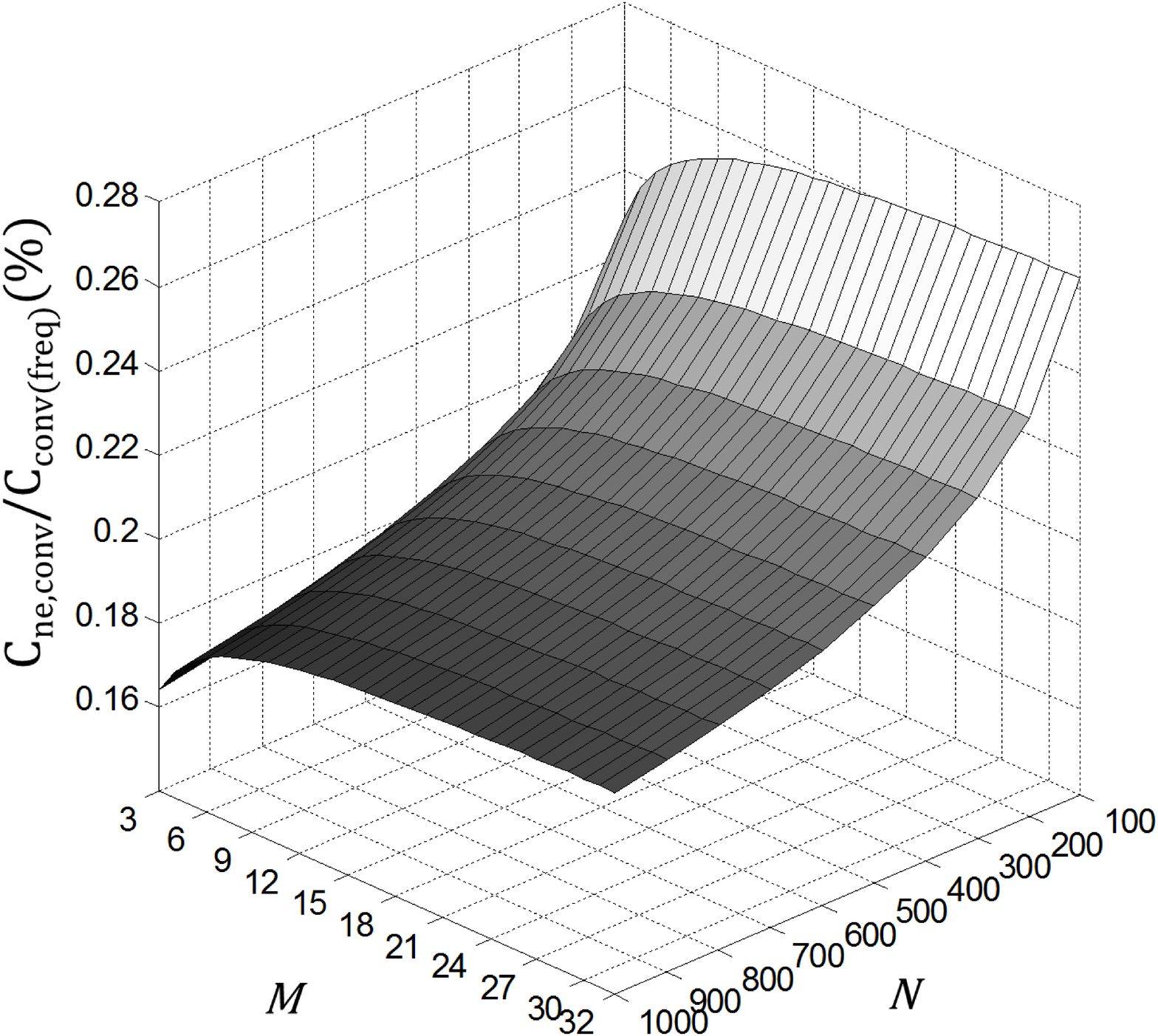}}
\par\end{centering}

\protect\caption{From left to right: Ratios of arithmetic operations for numerical
entanglement, extraction and result validation/recovery versus the
arithmetic operations of: generic matrix multiplication, time-domain
convolution and frequency-domain convolution, with $M$ the number
of streams (or the number of subblocks within a GEMM operation) and
$N$ the kernel size (the dimension of each subblock in a GEMM operation).
\label{fig:Entanglement-cost-vs.GEMM}}
\end{figure*}

\subsection{Complexity Analysis}

Consider $M$ input integer data streams, each comprising several
samples and consider that an LSB operation $\text{op}$ with kernel
$\mathbf{g}$ is performed in each stream. This is the case, for example,
under inner-products performed for GEMM or convolution/cross-correlation
between multiple input streams for similarity detection or filtering
applications or matrix-vector products in Lanczos iterations and iterative
methods \cite{golub1996matrix}. If the kernel $\mathbf{g}$ has substantially
smaller length than the length of each input stream, the effective
input stream size can be adjusted to the kernel length under overlap-save
or overlap-add operation in convolution and cross-correlation \cite{anam2012throughput}
and several (smaller) overlapping input blocks can be processed independently.
Similarly, block-major reordering is used in matrix products and transform
decompositions for increased memory efficiency \cite{anastasia2012throughput,goto2008anatomy,intel2007intel,andreopoulos2001local,andreopoulos2002new,andreopoulos2003high}.
Thus, in the remainder of this section we assume that $N$ expresses
\emph{both} the input data stream and kernel dimension under 32/64-bit
integer representation.

The operations count (additions/multiplications) for stream-by-stream
sum-of-products between a matrix comprising $M$ subblocks of $N\times N$
integers and a matrix kernel comprising $N\times N$ integers (see
\cite{anastasia2012throughput,goto2008anatomy,murray2008spread,chen2005fault}
for example instantiations within high-performance computing environments)
is: $\mathrm{C}_{\text{GEMM}}=MN^{3}$. For sesquilinear operations,
like convolution and cross-correlation of $M$ input integer data
streams (each comprising $N$ samples) with kernel $\mathbf{g}$ {[}see
Fig. \ref{fig:linear_operations_M_input_streams}(a){]}, depending
on the utilized realization, the number of operations can range from
$O\left(MN^{2}\right)$ for direct algorithms (e.g., time-domain convolution)
to $O\left(MN\log_{2}N\right)$ for fast algorithms (e.g., FFT-based
convolution) \cite{anam2012throughput}. For example, for convolution
or cross-correlation under these settings and an overlap-save realization
for consecutive block processing, the number of operations (additions/multiplications)
is \cite{anam2012throughput}: $\mathrm{C}_{\text{conv,time}}=4MN^{2}$
for time domain processing and $\mathrm{C}_{\text{conv,freq}}=M\left[\left(45N+15\right)\log_{2}\left(3N+1\right)+3N+1\right]$
for frequency-domain processing. 

As described in Section \ref{sec:From-Numerical-Packing-to-Entanglement},
numerical entanglement of $M$ input integer data streams (of $N$
samples each) requires $O\left(MN\right)$ operations for the entanglement,
extraction and validation (or recovery) per output sample. For example,
ignoring all arithmetic-shifting operations (which take a negligible
amount of time), based on the description of Section \ref{sec:From-Numerical-Packing-to-Entanglement}
the upper bound of the operations for numerical entanglement, extraction
and validation/recovery is: $C_{\text{ne,conv}}=2MN$. Similarly as
before, for the special case of the GEMM operation using $M$ subblocks
of $N\times N$ integers, the upper bound of the overhead of numerical
entanglement of all inputs is: $C_{\text{ne,GEMM}}=2MN^{2}$. We present
the percentile values obtained for $\frac{\mathrm{C}_{\text{ne,GEMM}}}{\mathrm{C}_{\text{GEMM}}}\times100\%$,
$\frac{\mathrm{C}_{\text{ne,conv}}}{\mathrm{C}_{\text{conv,time}}}\times100\%$
and $\frac{\mathrm{C}_{\text{ne,conv}}}{\mathrm{C}_{\text{conv,freq}}}\times100\%$
in Fig. \ref{fig:Entanglement-cost-vs.GEMM} for typical values of
$N$ and $M$. For sesquilinear operations, the overhead of numerical
entanglement, extraction and result validation/recovery in terms of
arithmetic operations is below $0.3\%$. Moreover, 
\begin{equation}
\lim_{N\rightarrow\infty}\frac{\mathrm{C}_{\text{ne,GEMM}}}{\mathrm{C}_{\text{GEMM}}}=\lim_{N\rightarrow\infty}\frac{\mathrm{C}_{\text{ne,conv}}}{\mathrm{C}_{\text{conv,time}}}=\lim_{N\rightarrow\infty}\frac{\mathrm{C}_{\text{ne,conv}}}{\mathrm{C}_{\text{conv,freq}}}=0,
\end{equation}
i.e., the overhead of the proposed approach approaches $0\%$ as the
dimension of the LSB processing increases. 

\begin{figure}[tbh]
\begin{centering}
\includegraphics[scale=0.13]{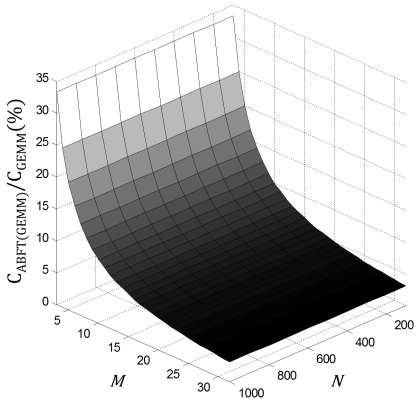}
\par\end{centering}

\protect\caption{Ratio of arithmetic operations for ABFT generation and result validation/recovery
versus the arithmetic operations of: generic matrix multiplication,
with $M$ the number of streams (i.e., GEMM subblocks) and $N$ the
kernel size (i.e., the dimension of GEMM subblocks). \label{fig:ECC-cost-vs.GEMM}}
\end{figure}

For comparison purposes, Fig. \ref{fig:ECC-cost-vs.GEMM} shows the
percentile overhead of ABFT methods {[}Fig. \ref{fig:linear_operations_M_input_streams}(b)
\cite{bosilca2009algorithm,chen2005fault,murray2008spread}{]} under:
\emph{(i)} the same range of values for $N$ and $M$ and \emph{(ii)}
the same fault tolerance capability%
\footnote{In order to keep our treatment generic, we consider as ABFT in GEMM
the method that generates an additional (i.e., checksum) subblock
for fault tolerance, instead of the row-column ABFT method of Huang
and Abraham \cite{huang1984algorithm}. However, our experiments present
a comparison of the proposed approach for GEMM against both approaches. %
}. Specifically, we examine the ratios: $\frac{\mathrm{C}_{\text{ABFT,GEMM}}}{\mathrm{C}_{\text{GEMM}}}\times100\%$,
$\frac{\mathrm{C}_{\text{ABFT,conv,time}}}{\mathrm{C}_{\text{conv,time}}}\times100\%$
and $\frac{\mathrm{C}_{\text{ABFT,conv,freq}}}{\mathrm{C}_{\text{conv,freq}}}\times100\%$,
where $\mathrm{C}_{\text{ABFT,GEMM}}=2MN^{2}+\frac{1}{M}\mathrm{C}_{\text{GEMM}}$,
$\mathrm{C}_{\text{ABFT,conv,time}}=2MN+\frac{1}{M}\mathrm{C}_{\text{conv,time}}$
and $\mathrm{C}_{\text{ABFT,conv,freq}}=2MN+\frac{1}{M}\mathrm{C}_{\text{conv,freq}}$
represent the overhead in terms of operations count (additions/multiplications)
for each case. Given that time-domain and frequency-domain convolution
exhibit the same percentile overhead as for the case of GEMM (with
variation that is limited to no more than $0.2\%$), Fig. \ref{fig:ECC-cost-vs.GEMM}
illustrates only the latter. As expected, the overhead of ABFT methods
converges to $\frac{1}{M}\times100\%$ as the dimension of the LSB
processing operations increases, i.e., 
\begin{eqnarray}
\lim_{N\rightarrow\infty}\frac{\mathrm{C}_{\text{ABFT,GEMM}}}{\mathrm{C}_{\text{GEMM}}} & = & \lim_{N\rightarrow\infty}\frac{\mathrm{C}_{\text{ABFT,conv,time}}}{\mathrm{C}_{\text{conv,time}}}\nonumber \\
 & = & \lim_{N\rightarrow\infty}\frac{\mathrm{C}_{\text{ABFT,conv,freq}}}{\mathrm{C}_{\text{conv,freq}}}\\
 & = & \frac{1}{M}.\nonumber 
\end{eqnarray}
Therefore, ABFT leads to substantial overhead (above $10\%$) when
high reliability is pursued, i.e., when $M\leq8$. Finally, even for
the low reliability regime (i.e., when $M>8$), Fig. \ref{fig:ECC-cost-vs.GEMM}
shows that ABFT can incur more than $4\%$ overhead in terms of arithmetic
operations. 

\subsection{Discussion}

The comparison between Fig. \ref{fig:Entanglement-cost-vs.GEMM} and
Fig. \ref{fig:ECC-cost-vs.GEMM} is illustrative for the capabilities
unleashed by the proposed highly-reliable numerical entanglement.
Evidently, in the proposed approach, the \emph{most-efficient operational
area} is the leftmost part of the plots, i.e. small values of $M$
and large values of $N$ (small-size grouping of long streams of high-complex
LSB operations). This area corresponds to the \emph{least-efficient
operational area} of ABFT. The comparison between the two figures
demonstrates that, for the same fault tolerance capability (e.g.,
detection of one transient fault in every three co-located outputs
or recovery from one fail-stop failure in three processors, which
corresponds to $M=3$), the overhead of the proposed approach is three
orders of magnitude smaller than that of ABFT. Conversely, the least-efficient
operational area for our approach is the rightmost part of the plots
of Fig. \ref{fig:Entanglement-cost-vs.GEMM} and Fig. \ref{fig:ECC-cost-vs.GEMM},
i.e. large values of $M$ and small values of $N$ (large-size grouping
of short streams of low-complex LSB operations). This area corresponds
to the \emph{most-efficient operational area} of high-redundant ABFT
methods. Nevertheless, the comparison between the two figures demonstrates
that, for the same fault tolerance capability (e.g., detection of
one transient fault in every 32 co-located outputs or recovery from
one fail-stop failure out of 32 processors, which corresponds to $M=32$),
the overhead of the proposed approach is still one to two orders of
magnitude less than that of ABFT. Overall, our approach is maximally
beneficial when high reliability is desired for complex LSB operations
with very low implementation overhead\textbf{.}

\section{Experimental Validation\label{sec:Experiments}}

\begin{figure*}[tbh]
\begin{centering}
\subfigure[$M=3$]{\includegraphics[scale=0.36]{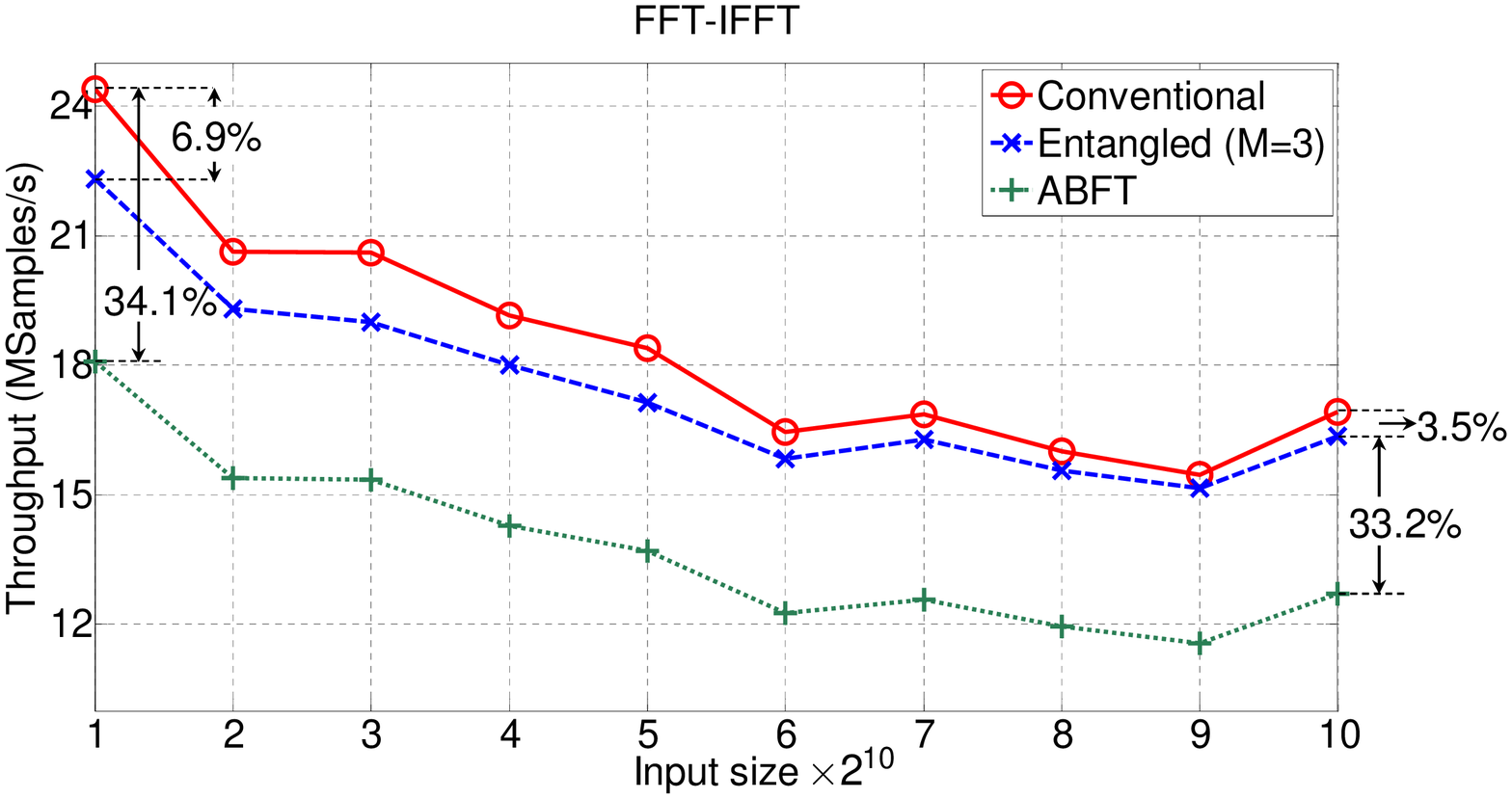}}
\subfigure[$M=8$]{\includegraphics[scale=0.36]{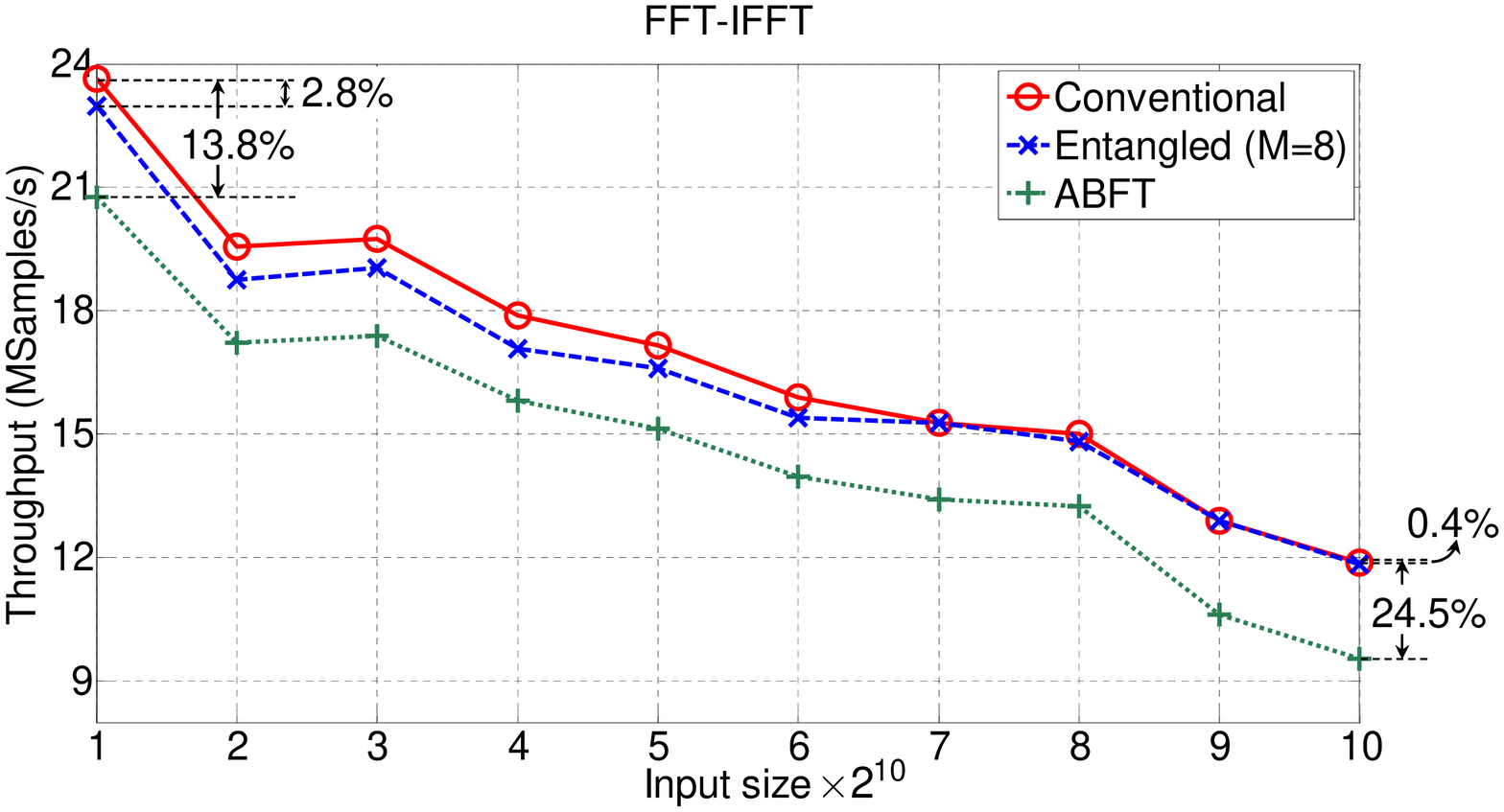}}
\par\end{centering}

\protect\caption{Throughput results for FFT--IFFT of $M$ integer streams. ``Conventional''
refers to conventional (fault-intolerant) FFT realization using FFTW
3.3.3 and it is used as a benchmark under (a) $M=3$; (b) $M=8$.
\label{fig:FFT-benchmark}}
\end{figure*}

\begin{figure*}[tbh]
\begin{centering}
\subfigure[$M=3$]{\includegraphics[scale=0.36]{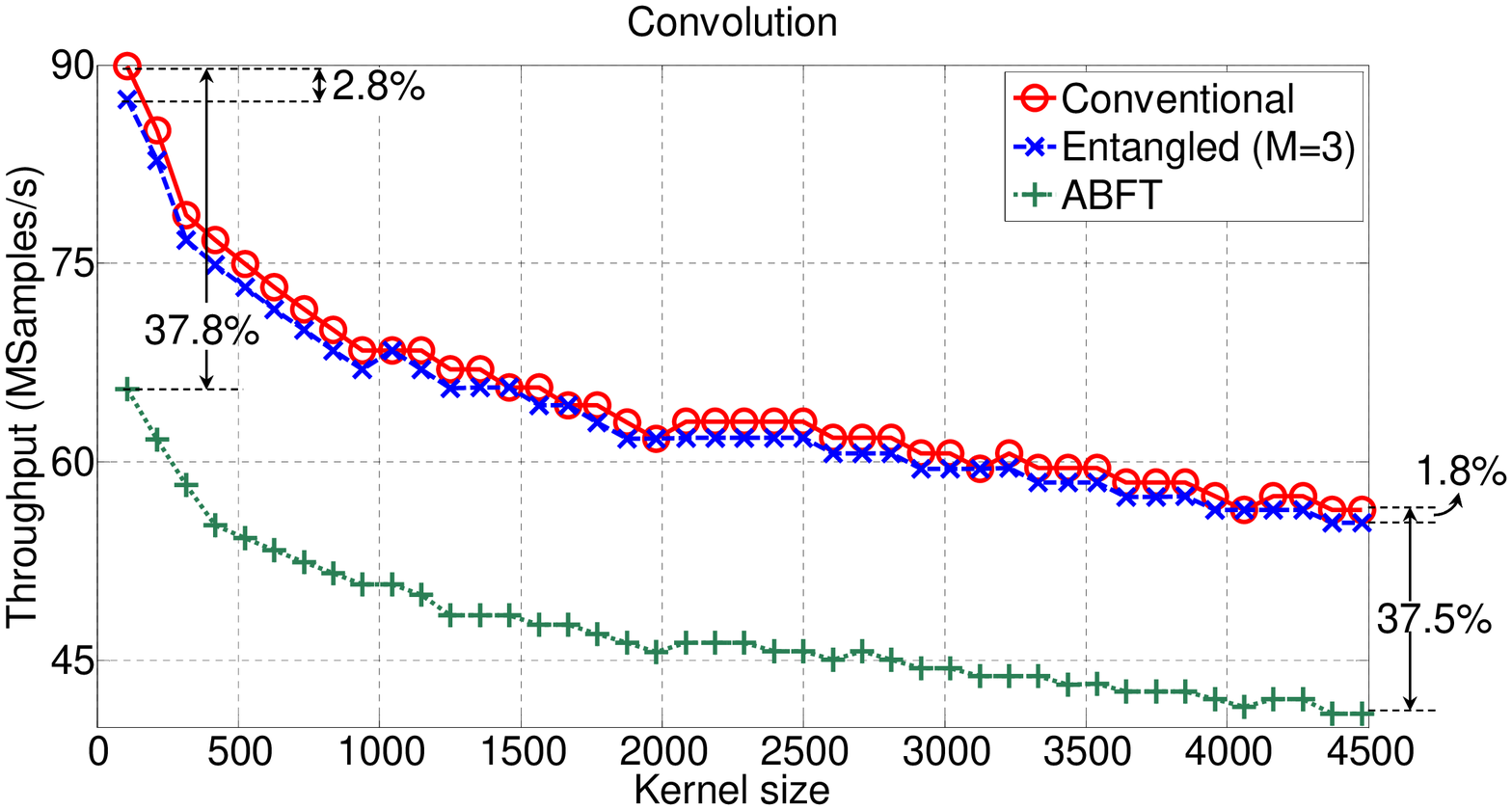}}
\subfigure[$M=8$]{\includegraphics[scale=0.36]{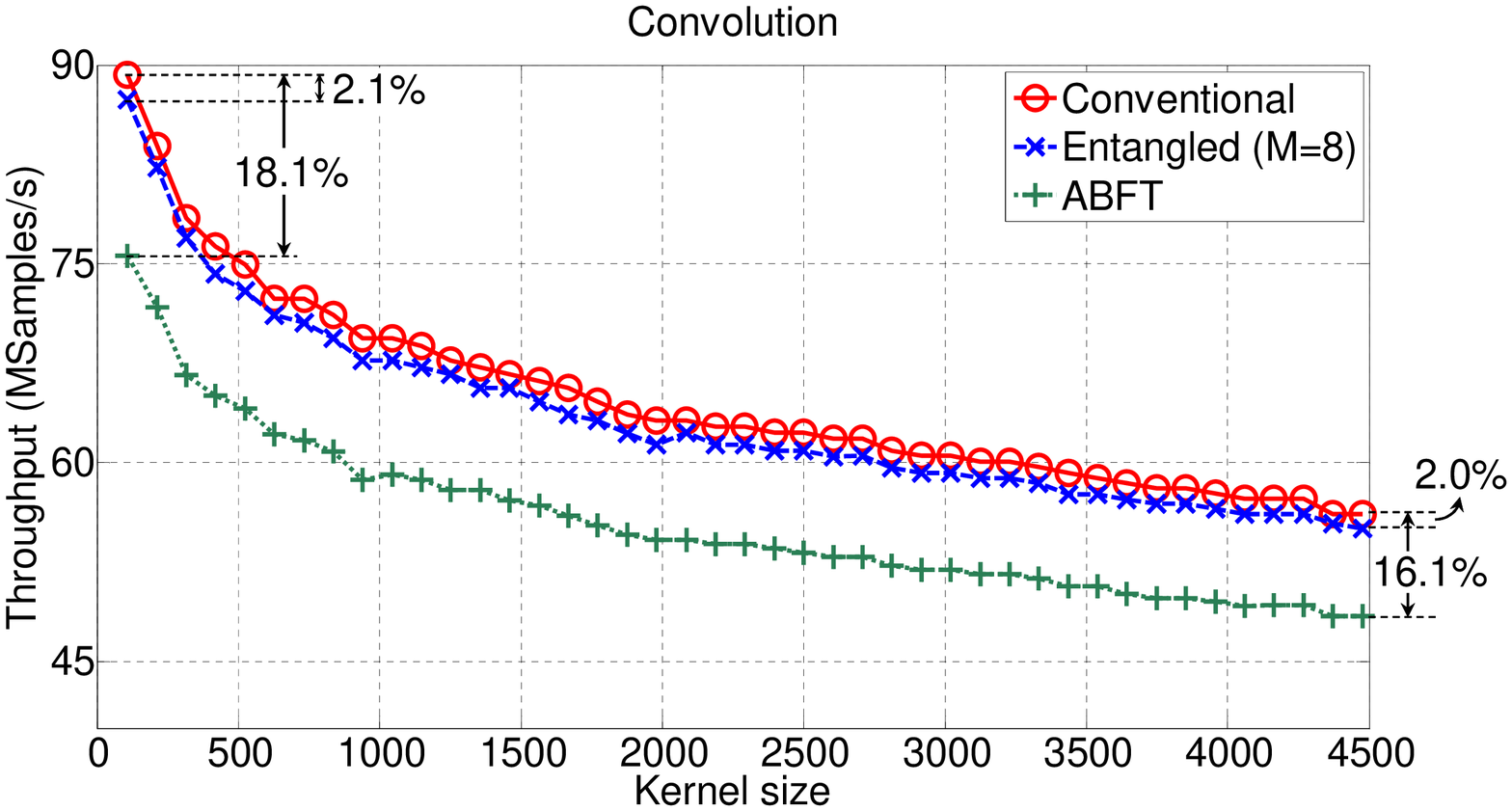}}
\par\end{centering}

\protect\caption{Throughout results for convolution of $M$ integer streams. ``Conventional''
refers to conventional (fault-intolerant) convolution realization
using Intel IPP 7.0 and it is used as a benchmark under (a) $M=3$;
(b) $M=8$. \label{fig:CONV-benchmark}}
\end{figure*}

\begin{figure*}[tbh]
\begin{centering}
\subfigure[$M=3$]{\includegraphics[scale=0.36]{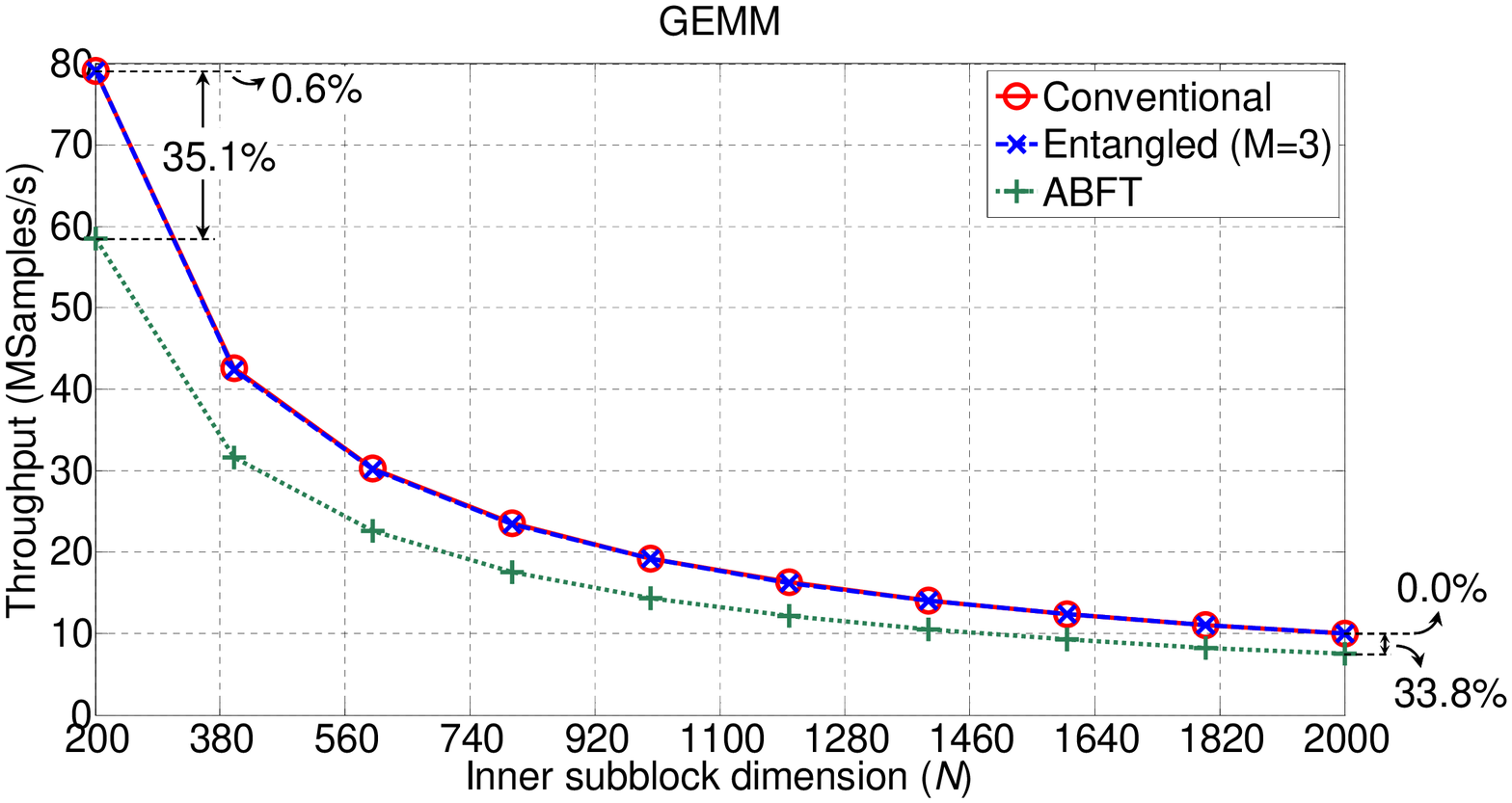}}
\subfigure[$M=8$]{\includegraphics[scale=0.36]{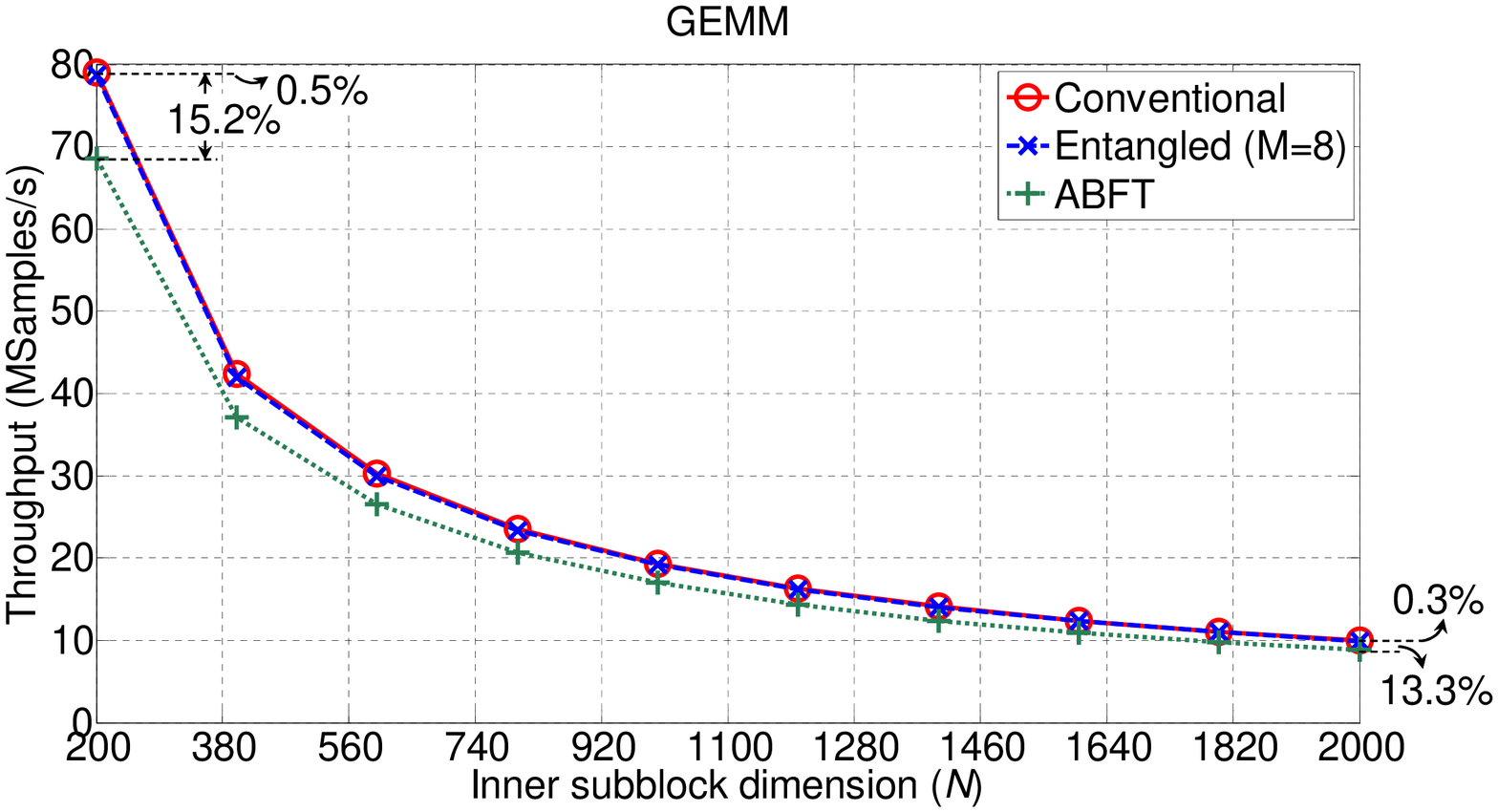}}
\par\end{centering}

\protect\caption{Throughput results for $M$ GEMM subblocks, each comprising a $2000\times N$
by $N\times1200$ integer matrix product. ``Conventional'' refers
to conventional (fault-intolerant) GEMM realization using Intel MKL
11.0 and it is used as a benchmark under (a) $M=3$; (b) $M=8$. \label{fig:GEMM-benchmark}}
\end{figure*}

All our results were obtained using an Intel Core i7-4700MQ 2.40GHz
processor (Haswell architecture with AVX2 support, Windows 8 64-bit
system, Microsoft Visual Studio 2013 compiler). Entanglement, disentanglement
and fault detection/recovery mechanisms were realized using the Intel
AVX2 SIMD instruction set for faster processing. For all cases, we
also present comparisons with ABFT-based fault tolerance, the checksum
elements of which were also generated using AVX2 SIMD instructions.

First, we consider the low-complexity case of basic frequency-domain
processing of $M$ streams (e.g., antialiasing, frequency-selective
systems, etc.) comprising: \emph{(i)} FFT of the input streams; \emph{(ii)}
attenuating a random subset of the resulting frequency components
of each stream to zero by weighting with the Fourier transform of
an integer frequency-selective filter; \emph{(iii)} inverse FFT (IFFT)
to obtain $M$ integer streams. We present results for input blocks
with dimension between $N\in\left\{ 2^{10},\ldots,10\times2^{10}\right\} $
and select two sizes for number of input streams, $M$, which represent
different fault tolerance capabilities and operational complexity
for the FFT--IFFT multi-stream realization. Fig. \ref{fig:FFT-benchmark}
presents representative results based on the FFTW 3.3.3 library \cite{frigo2012fftw}.
The throughput results are given in Mega-samples per second (Msamples/s).
The figure demonstrates that, under the same fault tolerance capability,
our approach leads to only $0.4\%$ to $6.9\%$ decrease in throughput,
while ABFT incurs throughput loss of $13.8\%$ to $34.1\%$.

We then extend this case by performing experiments with general convolution
operations of integer streams. We used Intel's Integrated Performance
Primitives (IPP) 7.0 \cite{taylor2003intel} convolution routine \texttt{ippsConv\_64f}
that can handle the dynamic range required under convolutions with
32-bit integer inputs. We experimented with: input size of $N_{\text{in}}=10^{6}$
samples, several kernel sizes between $N_{\text{kernel}}\in\left[100,\:4500\right]$
samples. Representative results are given in Fig. \ref{fig:CONV-benchmark}
under two settings for the number of input streams $M$. The results
demonstrate that the proposed approach substantially outperforms ABFT,
while allowing for same level of fault tolerance. In line with the
theoretical predictions of Fig. \ref{fig:Entanglement-cost-vs.GEMM}(b)
and Fig. \ref{fig:ECC-cost-vs.GEMM}(b), the decrease in throughput
for the proposed approach is only $1.8\%$ to $2.8\%$, while ABFT
incurs $16.1\%$ to $37.8\%$ throughput loss against the fault-intolerant
realization of convolution. 

Thirdly, considering generic matrix multiplication, out of several
sets of experiments performed, we present results for $M$ matrix
subblock products of $2000\times N$ by $N\times1200$ each, with
$N\in\left[200,\:2000\right]$. Fig. \ref{fig:GEMM-benchmark} presents
results for the decrease in throughput (in Mega samples per second)
against the conventional GEMM kernel realization based on the Intel
MKL GEMM subroutine \cite{intel2007intel}. In line with the theoretical
predictions of Fig. \ref{fig:Entanglement-cost-vs.GEMM}(a) and Fig.
\ref{fig:ECC-cost-vs.GEMM}(a) and for the same level of fault tolerance
capability, our approach is able to detect transient faults (or recover
from fail-stop failures) with only 0.03\% to 0.6\% decrease in throughput,
while ABFT requires $13.3\%$ to $35.1\%$ loss of throughput to perform
result validation within the final GEMM result. Therefore, ABFT is
2 to 3 orders of magnitude less efficient than the proposed approach. 

Finally, given that ABFT can be applied within individual GEMM subblocks
following the row-wise and column-wise checksum generation within
the input subblocks (as originally proposed by Huang and Abraham \cite{huang1984algorithm}),
we also carried out a comparison against such an ABFT framework for
GEMM (termed as ``ABFT RC-check''). The results demonstrated that,
for the same subblock sizes as for the experiments of Fig. \ref{fig:GEMM-benchmark},
the overhead of ABFT RC-check was between $3.5\%$ to $5.5\%$, which
is still 5 to 1000 times higher than the corresponding overhead of
the proposed approach. Importantly, while ABFT RC-check offers guaranteed
detection of transient faults in up to three outputs out of every
$2000\times1200$ outputs, the proposed approach offers guaranteed
detection/recovery of up to one fault out of every $M$ ($M\in\left[3,\:8\right]$
) outputs. Therefore, in this case, the significantly-increased efficiency
is also coupled with increased reliability for GEMM computations.

\section{Conclusion\label{sec:Conclusions}}

We propose a new approach for highly-reliable LSB processing of integer
data streams that is based on the novel concept of numerical entanglement.
Under $M$ input streams ($M\geq3$), the proposed approach provides
for: \emph{(i)} guaranteed detection of transient faults within any
single input/output stream; \emph{(ii)} guaranteed recovery from any
single fail-stop failure if each input stream is processed by a different
core; \emph{(iii)} complexity overhead that depends only on $M$ and
not on the complexity of the performed LSB operations, thus, quickly
becoming negligible as the complexity of the LSB operations increases.
These three features demonstrate that the proposed solution forms
a \emph{third family} of approaches for fault tolerance in data stream
processing (i.e., beyond ABFT and MR) and offers unique advantages,
summarized in Table \ref{tab:Summary-of-features}. As such, it is
envisaged that it will find usage in a multitude of systems that require
guaranteed reliability against transient or permanent faults in hardware
with very low implementation overhead.

\appendix{}

\subsection{Proof of Proposition 1}
\begin{IEEEproof}
Notice that \eqref{eq:integer_disentanglement} does not use $\delta_{0,n}$.
Therefore, full recovery of all outputs takes place even with the
loss of $\mathbf{\mathbf{\delta}}_{0}$. This occurs because, for
every $n$, $0\leq n<N_{\text{out}}$, $\delta_{1,n}$ and $\delta_{2,n}$
contain $\hat{d}_{0,n}$ and $\hat{d}_{2,n}$ {[}which are recovered
via \eqref{eq:integer_disentanglement}---this link is pictorially
illustrated in Fig. \ref{fig:Entanglement-with-M_equal_3}{]}. Therefore,
$\boldsymbol{\delta}_{1}$ and $\boldsymbol{\delta}_{2}$ suffice
to recover all three output streams $\hat{\boldsymbol{\mathbf{d}}}_{0}$,
$\hat{\mathbf{d}}_{1}$ and $\hat{\mathbf{d}}_{2}$. Since the entanglement
pattern is cyclically-symmetric, it is straightforward to rewrite
the disentanglement process of \eqref{eq:integer_disentanglement}
for recovery from any two out of $\boldsymbol{\delta}_{0}$, $\boldsymbol{\delta}_{1}$
and $\boldsymbol{\delta}_{2}$. 
\end{IEEEproof}

\subsection{Proof of Proposition 2}
\begin{IEEEproof}
Given that we can create any single entangled output stream from the
other two streams, we can apply this and compare the recreated entangled
output stream with the available one in order to check for occurrences
of transient faults. For example, we create $\boldsymbol{\hat{\delta}}_{0}$
starting from the recovered $\hat{\mathbf{d}}_{0}$ and $\hat{\mathbf{d}}_{2}$
(which were extracted solely from $\boldsymbol{\delta}_{1}$ and $\boldsymbol{\delta}_{2}$)
and then compare it with the original values of $\boldsymbol{\delta}_{0}$
to detect transient faults. Specifically, for every $n$, $0\leq n<N_{\text{out}}$,
if \\
\begin{eqnarray}
\delta_{0,n}-\left[\mathcal{S}_{l}\left\{ \hat{d}_{2,n}\right\} +\hat{d}_{0,n}\right] & \neq & 0\label{eq:unsigned_fault_check}
\end{eqnarray}
holds, then a transient fault occurred in the triplet of $\left\{ \delta_{0,n},\,\delta_{1,n},\,\delta_{2,n}\right\} $.
Since the recovered outputs $\hat{d}_{0,n}$ and $\hat{d}_{2,n}$
stem from the $3l+k$ least-significant bits of $d_{\text{temp}}$,
which in turn stem from the $2l+k$ bits of $\delta_{1,n}$ and $\delta_{2,n}$,
the check of \eqref{eq:unsigned_fault_check} includes all the bits
of $\left\{ \delta_{0,n},\,\delta_{1,n},\,\delta_{2,n}\right\} $.
Transient faults may still remain undetected if, and only if, there
exists a stream position $n$ ($0\leq n<N_{\text{out}}$) for which
\emph{two} or \emph{all three} out of $\left\{ \delta_{0,n},\,\delta_{1,n},\,\delta_{2,n}\right\} $
are corrupted in a manner that the check of \eqref{eq:unsigned_fault_check}
does not hold. Thus, we conclude that, for integer outputs with range
bounded by \eqref{eq:dynamic_range}, the check of \eqref{eq:unsigned_fault_check}
is \emph{necessary and sufficient} for the detection of any transient
fault in one of $\delta_{0,n}$, $\delta_{1,n}$, $\delta_{2,n}$
for all stream positions $n$, $0\leq n<N_{\text{out}}$.
\end{IEEEproof}

\subsection{Proof of Proposition 3}
\begin{IEEEproof}
For every output position $n$, $0\leq n<N_{\text{out}}$, we are
able to recover \emph{all results of all $M$ streams without using}
$\mathrm{\mathbf{\delta}}_{r,n}$ in \eqref{eq:integer_disentanglement_partial_M_general}--\eqref{eq:integer_disentanglement_M_general-1}.
Therefore, the proposed method is able to recover from a single fail-stop
failure in one of the $M$ entangled streams. 
\end{IEEEproof}

\subsection{Proof of Proposition 4}
\begin{IEEEproof}
Given that we can create any single entangled output stream from the
other $M-1$ streams, we can apply this and compare the created entangled
stream with the available one to check for the occurrences of transient
faults. For example, if we apply this for the $r$th entangled output
stream, $\boldsymbol{\delta}_{r}$, ($1\leq r\leq M$, $0\leq n<N_{\text{out}}$):
\\
\begin{equation}
\mathrm{\mathbf{\delta}}_{r,n}-\left[\hat{d}_{\left(r+M\right)\text{mod}M,n}+\mathcal{S}_{l}\left\{ \hat{d}_{\left(r+M-1\right)\text{mod}M,n}\right\} \right]\neq0\label{eq:unsigned_fault_check_M_general}
\end{equation}
then, if \eqref{eq:unsigned_fault_check_M_general} holds, a fault
occurred in one of the $M$ entanglements at stream position $n$.
Since the recovered outputs $\hat{d}_{\left(r+m\right)\text{mod}M,n}$
and $\hat{d}_{\left(r+M-1\right)\text{mod}M,n}$ stem from the $2\left(M-1\right)l+k$
least-significant bits of $d_{\text{temp}}$, which in turn stem from
the $\left(M-1\right)l+k$ bits of $\left\{ \delta_{0,n},\,\ldots,\,\delta_{r-1,n},\,\delta_{r+1,n},\ldots,\,\delta_{M-1,n}\right\} $,
the check of \eqref{eq:unsigned_fault_check_M_general} includes all
the bits of $\left\{ \delta_{0,n},\,\ldots,\,\delta_{M-1,n}\right\} $.
Transient faults may still remain undetected if, and only if, there
exists a stream position $n$ ($0\leq n<N_{\text{out}}$) for which
two or more out of $\left\{ \delta_{0,n},\,\ldots,\,\delta_{M-1,n}\right\} $
are corrupted in a manner that the check of \eqref{eq:unsigned_fault_check_M_general}
does not hold. Thus, we conclude that, for integer outputs with range
bounded by \eqref{eq:dynamic_range}, the check of \eqref{eq:unsigned_fault_check_M_general}
is \emph{necessary and sufficient} for the detection of any transient
fault in one of $\left\{ \delta_{0,n},\,\ldots,\,\delta_{M-1,n}\right\} $
for all stream positions $n$, $0\leq n<N_{\text{out}}$.
\end{IEEEproof}
\bibliographystyle{IEEEbib}
\bibliography{literatur}

\end{document}